\documentclass[useAMS,usenatbib,usegraphicx]{mn2e}
\usepackage{amsmath,amsfonts,amssymb}

\newcommand{\lya}{Ly$\alpha$}
\newcommand{\di}{\textrm{d}}
\newcommand{\h}{\textrm{H}}
\newcommand{\hi}{\textrm{H {\tiny I}}}
\newcommand{\hii}{\textrm{H \tiny{II}}}

\title[Galactic Winds]{How do galactic winds affect the \lya\ forest?}

\author[Bertone \& White]
{Serena Bertone$^{1,2}$\thanks{E-mail: s.bertone@sussex.ac.uk}
and Simon D.M. White$^{2}$ \\
$^{1}$Astronomy Centre, University of Sussex, Brighton BN1 9QH, United Kingdom \\
$^{2}$Max Planck Institut f\"ur Astrophysik, Karl Schwarzschild Str. 1,
85741 Garching bei M\"unchen, Germany}

\begin{document}

\date{Submitted to MNRAS}
\pagerange{\pageref{firstpage}--\pageref{lastpage}} \pubyear{2005}
\maketitle
\label{firstpage}

\begin{abstract}
We investigate the effect of galactic winds on the \lya\ forest in
cosmological simulations of structure and galaxy formation. We combine
high resolution N-body simulations of the evolution of the dark matter
with a semi-analytic model for the formation and evolution of galaxies
which includes detailed prescriptions for the long-term evolution of
galactic winds. This model is the first to describe the
evolution of outflows as a two-phase process (an adiabatic bubble
followed by a momentum--driven shell) and to include metal--dependent
cooling of the outflowing material. We find that the main
statistical properties of the \lya\ forest, namely the flux power
spectrum $P(k)$ and the flux probability distribution function (PDF),
are not significantly affected by winds and so do not significantly
constrain wind models. Winds around galaxies do, however, produce
detectable signatures in the forest, in particular, increased flux
transmissivity inside hot bubbles, and narrow, saturated absorption
lines caused by dense cooled shells.  We find that the \lya\ flux
transmissivity is highly enhanced near strongly wind--blowing
galaxies, almost half of all high-redshift galaxies in our sample, in
agreement with the results of Adelberger et al. (2005).  Finally, we
propose a new method to identify absorption lines potentially due to
wind shells in the \lya\ forest: we calculate the abundance of
saturated regions in spectra as a function of region width and we find
that the number with widths smaller than about 1 \AA\ at $z=3$ and 0.6
\AA\ at $z=2$ may be more than doubled. This should be detectable in
real spectra.

\end{abstract}


\section{Introduction}
\label{intro}

Extensive observations of QSO spectra have provided a wealth
of information on properties of the intergalactic medium (IGM).

The \lya\ forest provides a measure of the power
spectrum of density fluctuations $P_{\textrm{m}}(k)$ on smaller scales
than those accessible by Cosmic Microwave Background (CMB)
observations (\citealt{spergel}, \citealt{vielcmb}) or large-scale
structure surveys (e.g. \citealt{tegmark}, \citealt{viana}, \citealt{colless}).
The flux power spectrum $P(k)$ of the \lya\ forest is directly related
to the power spectrum of density fluctuations $P_{\textrm{m}}(k)$,
thanks to the existence of a direct connection between the \lya\
absorption and the underlying density and velocity fields
(\citealt{viel2004}, \citealt{croft}, \citealt{hui}, \citealt{croft1998}).  The
flux power spectrum $P(k)$ has been estimated at redshifts $1 < z < 4$
(\citealt{kim}, \citealt{mcsdss}, \citealt{croft}, \citealt{mcd},
\citealt{croft1998}) for $0.001 < k < 0.1$ s km$^{-1}$, corresponding to
scales up to about 50 $h^{-1}$ comoving Mpc. Given that the data sets
used by different groups were taken with different instruments
(e.g. UVES, HIRES, LRIS) on different telescopes (e.g. VLT, Keck, SDSS
Apache Point) and at different resolution and signal--to--noise ratio,
the agreement between the results is good.

When using the matter power spectrum $P_{\textrm{m}}(k)$ from the
\lya\ forest to estimate cosmological parameters in conjunction with
CMB and large-scale structure data sets, it is important to know
precisely how $P(k)$ is affected by sources of systematic errors and
non--cosmological distortions. For example, the estimation of the
unabsorbed continuum level in a QSO spectrum is extremely sensitive to
fluctuations in the cosmic UV background radiation, which is poorly
constrained, and this affects both the amplitude and slope of $P(k)$
(\citealt{lidz}, \citealt{mc2005}, \citealt{viel}, \citealt{hui}). High column
density absorbers with damping wings and signatures of galactic winds
may produce ``local'' distortions in the \lya\ forest which introduce
deviations in the power spectrum at smaller scales (\citealt{fang},
\citealt{mc2005}, \citealt{viel}). The presence of metal lines blended with
the \lya\ forest affects the flux power spectrum at very small scales
\citep{kim}.

Adelberger et al. (2003) jointly analysed high resolution quasar
spectra and spectra of Lyman break galaxies (LBG) near the quasar
lines-of-sight. Their earliest results indicated that the \lya\
continuum emitted by the background quasars can be almost completely
transmitted by the intergalactic medium (IGM) near LBGs.  
Numerical simulations (\citealt{vincent}, \citealt{maselli}, \citealt{juna})
have been unable to reproduce this result, predicting instead a decrease
in the \lya\ flux transmissivity with decreasing impact parameter (the
impact parameter is defined as the minimum distance between a
galaxy and the quasar line-of-sight).  When new results derived from a larger
observational sample were released (Adelberger et al. 2005, hereafter
A05), the improved statistics showed that the mean
transmitted flux around LBG does indeed decrease with
decreasing impact parameter, as predicted by the simulations, but
about a third of the galaxies in the full sample still show little or
no \lya\ absorption in their proximity. This suggests that the earlier
result of \citet{adelb} was affected by small number statistics,
although even a relatively small fraction of LBG's
``bubbles'' in absorption appears inconsistent with published simulations.

In this work, we use a semi--analytic model for the long--term
evolution of galactic winds by Bertone, Stoehr \& White (2005, BSW05 hereafter) to investigate the effect of galactic winds
on the \lya\ forest. In particular, we calculate the flux probability
distribution function (PDF), the flux power spectrum $P(k)$ and the
\lya\ flux transmissivity near high redshift galaxies, and
we compare the numerical results with the relevant observations. We
consider three different models, a ``no wind'' model and two wind
models corresponding to qualitatively different scenarios, and we try
to determine which model better reproduces the observations. In
addition, the comparison of the ``no wind'' model with the two wind
models highlights the effects of galactic winds on the \lya\ forest.  
To do this, we extract artificial spectra from
the ``M3'' set of high resolution N--body simulations \citep{felix},
which combines a large dynamical range with a high mass resolution
. The semi--analytic wind model of BSW05 includes a new
description of the dynamics of the winds as a two--phase process: a
first phase of adiabatic evolution during which the gas outflowing
from galaxies is hot, and a second snowplough phase in which the wind
material cools down and accumulates in a thin shell pushed by the
momentum of the wind. The efficiency of winds in the simulations is
set by a parameter ($\varepsilon$, see Subsection \ref{winds}) that
determines the amount of mass swept up by the expanding gas.

The paper is organised as follows. In Section \ref{two} we briefly
describe our set of simulations and the semi--analytic model for winds
of BSW05. Section \ref{loes} presents our method to extract
artifical spectra from the simulations and to include the
contributions of winds. Detailed descriptions of our calculation of
the hydrogen ionisation state and of the flux spectrum are given in
Appendix \ref{ion} and Appendix \ref{optical} respectively.  Results
are presented in Section \ref{fpdf} for the flux PDF and in Section
\ref{fpower} for the power spectrum. In Section \ref{fadel} we
present our results for the flux transmissivity around LBG and we
discuss the observations of A05. In Section \ref{gap} we
present a new method to test statistically for the signature of winds
in spectra. Section \ref{conclu} contains our conclusions.

\section{Semi--analytic simulations of galactic winds}
\label{two}

Since we want to investigate the effects of outflows in their proper
cosmological context, we opt for a combination of N--body simulations
for the evolution of the dark matter density field and a
semi--analytic model for the formation and evolution of galaxies. This
choice gives us the possibility to efficiently combine a high
resolution in mass with a large simulated volume, as is necessary
to study the effects of feedback from galaxies with a wide range in
stellar mass. The large volume allows us to achieve a good statistics
when investigating the effect of different galaxy populations on their
surroundings.  Our simulations assume a $\Lambda$CDM cosmology with
matter density $\Omega_m =0.3$, dark energy density $\Omega_{\Lambda}
=0.7$, Hubble constant $h=0.7$, primordial spectral index $n=1$ and
normalisation $\sigma_8 = 0.9$.

In the following subsections we briefly describe our set of numerical
simulations (Subsection \ref{sims}) and our semi--analytic model for
the physics of galactic winds (Subsection \ref{winds}). However, for a
more detailed description of the wind model, we refer the interested
reader to BSW05.

\subsection{The M3 simulations}
\label{sims}

We use the ``M3'' high-resolution N--body simulation of \citet{felix}.
M3 is a resimulation at higher resolution of an approximately
spherical region of the universe with average density close to the
cosmic mean and diameter 52 $h^{-1}$ Mpc. The particle mass in the
high resolution region is $1.7\cdot 10^8 h^{-1}$ M$_{\sun}$ and the
number of particles about $7\cdot 10^7$.  The simulations were
performed using the parallel treecode {\small GADGET I}
\citep{gadget} and 52 simulation outputs were stored between $z=20$
and $z=0$.

The formation and evolution of galaxies is modelled with the
semi--analytic technique of \citet{semi}. Merging trees extracted from
the simulations are used to follow the galaxy population in time,
while simple prescriptions for gas cooling, star formation and galaxy
merging model the processes involving the baryonic component of the
galaxies.  At $z=3$ a total of about four hundred thousand galaxies
are identified and about three hundred and fifty thousand are present
at $z=0$.  About half of the galaxies at $z=0$ are field galaxies,
while the rest are in groups and poor clusters. The two largest
clusters have a total mass of about $10^{14} h^{-1}$ M$_{\sun}$.  The
recipes for the evolution of winds are implemented on top of this
pre--existing scheme.

\subsection{The wind model}
\label{winds}

BSW05 introduced new prescriptions for the evolution of galactic winds
in the semi--analytic model for galaxy formation of \citet{semi}.  The
most innovative feature of their prescriptions is a description of
the dynamics of outflows as a two--phase process: a pressure--driven
adiabatic expansion followed by a momentum--driven snowplough.

Previous simulations of galactic winds in a cosmological context usually assume that winds are either pressure--driven or momentum--driven.
The wind model of \citet{sh} mimics the behaviour of an adiabatic bubble expanding into the surrounding medium. However, since metal cooling is neglected, \citet{sh} predict too high temperatures for the ejected metals and they are unable to reproduce the observed ${\textrm{C \small{IV}}}$ absorption in QSO spectra \citep{aguirre2005}.
\citet{tv} find that winds have little effect on the ${\textrm{H \small{I}}}$ absorption statistics, because the hot bubbles mostly escape into the voids.
Semi--analytic models of galaxy formation that include winds usually describe outflows as momentum--driven snowploughs \citep{aguirre}.

We do not resolve the first phases of the wind evolution, when a
superbubble sweeps through the ISM of a galaxy. Instead, we follow the
long--term evolution of winds once they have escaped the visible
regions of galaxies.  We make the simplifying assumption of spherical
symmetry for the wind evolution.  This may seem a rough approximation
at $z\sim 0$, where outflows are mostly bipolar, but it is a good
assumption at higher redshifts, where outflows seem to have a more
spherical geometry (\citealt{shapley}, \citealt{rupke}).

Galactic winds are modelled as uniform pressure--driven bubbles of hot
gas emerging from star--forming galaxies. The assumption that winds are
adiabatic at blow--out is motivated by observations of galaxies in the
local universe. For example, \citet{hoopes} and \citet{ses} observe no
energy losses in M82 through radiative cooling of the coronal ($T \sim
10^{5.5}$ K) and the hot ($T \sim 10^7$ K) phases of the wind,
supporting the idea that the early evolution of this wind is nearly
adiabatic.  The adiabatic phase of the wind evolution is described by
the equation for the conservation of energy $E$ \citep{omk}:
\begin{eqnarray}\label{energy}
   \frac{\di E}{\di t} & = & \frac{1}{2}\dot{M}_{\textrm{w}} v_{\textrm{w}} ^2 
   + \varepsilon 4\pi R^2 \cdot \nonumber \\
   & & \left\{ \left[ \frac{1}{2}\rho_{\textrm{o}} v_{\textrm{o}} ^2 + 
   u_{\textrm{o}} - \rho_{\textrm{o}} \frac{GM_{\textrm{h}}}{R} \right] \left( 
   v_{\textrm{s}} - v_{\textrm{o}} \right) - v_{\textrm{o}} P_{\textrm{o}} 
   \right\} ,
\end{eqnarray}
where $R$ and $v_{\textrm{s}}$ are the radius and the velocity of the
shock, $\dot{M}_{\textrm{w}}$ and $v_{\textrm{w}}$ the mass outflow
rate and the outflow velocity of the wind, $\rho_{\textrm{o}}$,
$P_{\textrm{o}}$ and $v_{\textrm{o}}$ the density, the pressure and
the outward velocity of the surrounding medium and $M_{\textrm{h}}$
the total mass internal to the shock radius.  The entrainment fraction
parameter $\varepsilon$ defines the fraction of mass that the wind
sweeps up while crossing the ambient medium.  Here, we call
``entrained'' gas the ambient gas which has mixed into the hot bubble
phase either through turbulent mixing of shocked diffuse ambient gas
or by evaporation and ablation of the filament gas, most of which ($1
-\varepsilon $) continues falling onto the galaxy. The latter process
is similar to the loading of ISM mass onto stellar winds and supernova
blastwaves.

The adiabatic phase is terminated when the loss of energy by radiation
becomes substantial and most of the energy transferred to the
swept--up gas is radiated away. When the cooling time of the hot
bubble becomes shorter than the age of the wind, a thin shell of
cooled gas forms near the bubble's outer boundary and continues to
expand pushed by the momentum of the wind.  The snowplough phase is
described by the equation for the conservation of momentum:
\begin{eqnarray}\label{momentum}
   \frac{\di}{\di t}\left( m v_{\textrm{s}} \right)
   & = & \dot{M}_{\textrm{w}} \left( v_{\textrm{w}} - v_{\textrm{s}} \right) - 
   \frac{GM_{\textrm{h}}}{R^2} m - \nonumber \\
   & & \varepsilon 4\pi R^2 \left[ P_{\textrm{o}} + \rho_{\textrm{o}} 
   v_{\textrm{o}} \left( v_{\textrm{s}} - v_{\textrm{o}}\right) \right],
\end{eqnarray}
where $m$ is the mass of the shell.

In our simulations both bursts of star formation and quiescent star
formation can power winds, since we do not explicitly 
restrict the star formation rate required for a galaxy to blow a
wind.  It is not possible to predict \emph{a priori} when a wind will
escape the gravitational pull of a galaxy, since its evolution and its
final fate are linked to several factors, like the star formation and
the mass accretion history of the galaxy, the potential well of the
dark matter halo in which it expands, the amount of mass accreted from
the wind and from the IGM and so on.

In general, a wind receives energy from the starburst and is slowed
down by the gravitational attraction of the central galaxy and by the
ram pressure of the ambient medium. Thermal pressure effects are
included consistently inside cluster haloes, but are neglected in the
IGM.  If the entrained mass is small (e.g. $\varepsilon \leq 0.1$), a
large fraction of the bubble or shell mass consists of supernova
ejecta and shocked ISM, which flow out from the galaxy with a velocity
often much larger than the escape velocity of the halo.  Since little
energy or momentum has to be spent by the wind to accelerate the
entrained mass, the shock velocity is less sensitive to energy losses
by pressure and gravity. Such a wind is thus more likely to break free
from the halo than more mass-loaded winds.  When mass-loading is
substantial (e.g. $\varepsilon =0.3$), a significant part of the wind
energy is consumed to accelerate the entrained gas and the expansion
slows down considerably.  If the amount of swept--up mass is large
compared to the mass initially ejected, the shock velocity may drop
below the escape velocity from the galaxy and the wind collapses back
onto the galaxy.

At high redshift winds tend to be mostly momentum--driven, while at
lower redshifts bubbles are much more likely to remain adiabatic.
This is partly due to the higher mean density of the Universe at high
$z$ and partly to a lower energy input from star formation, which
results in lower bubble temperatures and shorter cooling times
immediately after blowout.  The transition from pressure--driven
bubbles to momentum--driven shells may be a first hint that a wind is
not powerful enough to escape the galaxy's attraction. In fact,
pressure--driven winds are overall more likely to escape galaxies than
momentum--driven ones.

\section{Simulating the \lya\ forest}
\label{loes}

In this section we show how we extract sets of artificial spectra from
our simulated region. Since we use a semi--analytic model associated
to N--body simulations, we do not have \emph{a priori} all the
information about the gas properties provided by a gas-dynamical
simulation. We therefore need to estimate quantities like the gas
temperature and density, which are necessary to build the spectra, by
applying approximate prescriptions. We show how in Subsection
\ref{gas}. In Subsection \ref{displace} we describe how we include the
effects of winds in the calculation of the spectra.

We extract lines-of-sight (LOS, hereafter) from our simulated region
following the numerical scheme of \citet{tom}. We describe our
assumptions about the ionisation state of the gas along the LOS in
Appendix \ref{ion}. In Appendix \ref{optical} we summarize our
prescriptions for the integration of the relevant quantities along the
LOS and for the calculation of the optical depth and the flux
spectrum. Each spectrum is normalized to the mean observed flux of
\citet{kim02}.  Gaussian noise is added to the simulated spectra assuming
a fixed signal--to--noise ratio of 50, typical of {\small HIRES}
spectra. Gaussian read--out noise with variance $0.004^2$ is also
added to each pixel.

We extract synthetic spectra from three different sets of simulations:
1) a ``no wind'' model, in which we do not include the effects of
galactic winds; 2) a wind model with low mass loading efficiency and
entrainment fraction parameter $\varepsilon =0.1$ (hereafter ``e01'');
3) a wind model with high mass loading efficiency and entrainment
fraction parameter $\varepsilon =0.3$ (hereafter ``e03''). For each
set of simulations we extract spectra at $z\sim 3$ and $z\sim 2$.

For each wind model, we then extract two sets of spectra from our
simulated region: i) one set along random directions and ii) one set
imposing that the LOS intercept specially selected galaxies at random
distances in the interval (0, 10) $h^{-1}$ Mpc comoving. We use the
first sample of spectra to calculate the flux probability distribution
function (PDF) and the power spectrum $P(k)$ and the second to
investigate the effects of winds on the surroundings of galaxies. The
second set of spectra was selected to match the observations of
A05 as accurately as possible. In table 3 of
A05 are listed the properties of the galaxies observed
by {\small NIRSPEC} with impact parameters (e.g. distance galaxy--LOS)
smaller than 1 $h^{-1}$ Mpc comoving.  According to these data, we
select galaxies in our simulation which have: 1) stellar masses larger
than $10^9$ M$_{\sun}$; 2) SFR larger than 1 M$_{\sun}$/yr: these two
values are the minimum stellar mass and the minimum SFR estimated by
A05. In total we extract a sample of about 4400 LOS at
$z\sim 3$ and about 3300 LOS at $z\sim 2$: one LOS for every galaxy
with the desired properties. We make no requirements about the
presence of winds, but we do require that both the selected galaxy and
the LOS fall inside our high resolution region.

The set of random spectra has been selected assuming that each LOS
passes at a distance of less than half the radius of the high
resolution region from the centre of mass of the simulation
$\mathbf{x_{\textrm{c}}}$, that is $d\left( \textrm{LOS},
\mathbf{x_{\textrm{c}}} \right) < 13$ $h^{-1}$ Mpc. This is because we
want to extract simulated spectra of minimum length 45 $h^{-1}$ Mpc
comoving.

\subsection{Density and temperature of the gas}
\label{gas}

\begin{figure}
\centering
\includegraphics[width=8.8cm]{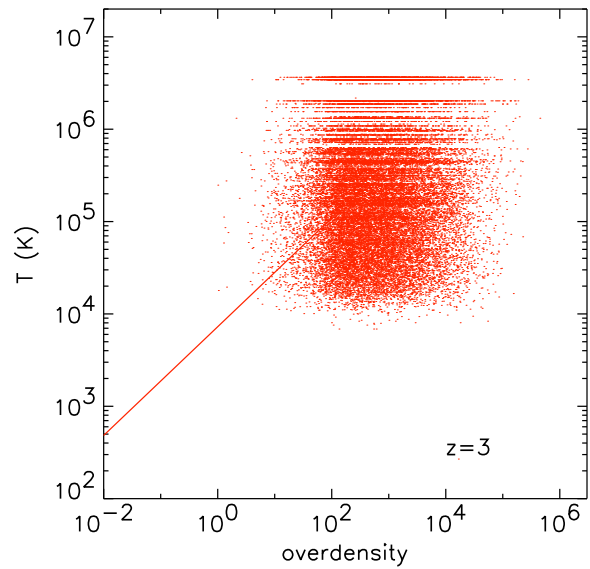}
\caption{The temperature of the gas in M3 at $z=3$. Gas that obeys
the equation of state in eq. \ref{state} piles up on the straight
line, while the shocked gas in haloes is spread out in a cloud--like
region.}
\label{igmtemp}
\end{figure}

The gas density in our simulated region is assumed to follow the dark
matter density distribution. This is not a very satisfactory
description for large matter overdensities, like galaxies or galaxy
groups, but it is sufficiently accurate for the gas in low density
regions, which represents about 70\% of the mass at $z\sim 3$ and
which is responsible for most of the \lya\ forest. The density of the dark
matter $\rho_{\textrm{DM}}$ is calculated by SPH interpolation over 26 neighbours and the
density of the gas $\rho_{\textrm{gas}}$ is recovered by assuming that
$\rho_{\textrm{gas}} = f_{\textrm{baryon}} \cdot \rho_{\textrm{DM}}$,
with $f_{\textrm{baryon}}$ the global baryon fraction.

The temperature of the ``gas'' particles is calculated separately for
high density particles bound to haloes and low density unbound
particles. This double treatment is necessary since the physical
conditions of the gas in the two cases are different. The
intergalactic gas is mostly unshocked, cools adiabatically and is
photoionised by the UV background radiation. On the other hand, the
gas in haloes is shock--heated because of the structure formation
process and its temperature is determined by the gravitational
potential of the dark matter halo in which it resides.  We therefore
divide our particles in two subgroups, the ``halo'' particles and the
``IGM'' particles, and calculate their temperatures according to
different prescriptions.

For ``halo'' particles we mean all those particles which belong to
bound structures and we assume that the temperature of these particles
equals the virial temperature of the halo \citep{wf}:
\begin{equation}\label{viritemp}
T_{\textrm{vir}} = \frac{\mu m_{\textrm p} V_{\textrm c} ^2}{2k_{\textrm B}} = 35.9 \left[ \frac{V_{\textrm c}}{\textrm{km
s}^{-1}} \right]^2 \textrm{ K},
\end{equation}
where $V_{\textrm c}$ is the circular velocity of the halo, $\mu$ is
the mean molecular weight of the gas, $m_{\textrm p}$ is the proton
mass, $\mu m_{\textrm p}$ the mean particle mass and $k_{\textrm B}$
the Boltzmann constant.

The low density ``IGM'' particles represent the diffuse intergalactic
medium and usually have overdensities $\delta\lesssim 10 - 100$.  For
these particles, we calculate the temperature from the equation of
state of \citet{huignedin}, which approximates a power law:
\begin{equation}\label{state}
T = T_{\textrm o} \left( 1+ \delta\right)^{\gamma -1},
\end{equation}
where $\gamma -1 \sim 1/1.7$. $T_{\textrm o}$ is the temperature
of the IGM at the mean density and can be recovered under the
assumption that the gas is in photoionisation equilibrium
(\citealt{huignedin}, \citealt{joop}). In this work, we use the results of
\citet{bolton} to define $T_{\textrm o}$ as a function of redshift.

In figure \ref{igmtemp} we show the temperature of the gas in M3 at
$z=3$. The low density gas obeying the equation of state (\ref{state})
is distributed along a straight line, while the high temperature gas
in haloes is spread into a cloud--like region. The horizontal lines
crossing the cloud are due to gas with different densities residing in
the same halo, which is assumed to behave as an isothermal gas sphere.
Our prescriptions for the gas temperature are unable to reproduce the
cold neutral gas in galaxies and galaxy haloes. This will partly
affect our results and we will show how in the following Sections.

\subsection{The matter displaced by winds}
\label{displace}

To include the effects of winds in the calculation of a spectrum, we
proceed in the following way.  As seen in Section \ref{winds}, winds
can be described as bubbles of hot gas or as thin shells of cold
material accumulating at the edge of a cooled cavity.  These two
phases of the wind evolution produce different conditions for the
material displaced by winds and a separate treatment of bubbles and
shells is therefore needed.

Since winds modify the spatial distribution of the gas near galaxies,
we need to identify how much mass is displaced, which particles can be
associated with the displaced matter and where this matter moves. All
the particles residing outside winds remain unaffected.  To do this,
we identify all the gas particles positioned inside winds. We then
calculate the total amount of mass $M_{\textrm{in}}$ inside winds as
predicted by the semi--analytic model: this is equal to the sum of the
stellar and cold gas mass of galaxies plus the halo or IGM mass inside
the wind radius.  The total mass contained in bubbles, shells and
cavities is $M_{\textrm{out}} = \sum \left( M_{\textrm{w}} +
m_{\textrm{e}} \right)$, with $M_{\textrm{w}}$ the mass of the wind
ejecta and $m_{\textrm{e}}$ the swept--up mass in each wind. We
finally define $F = M_{\textrm{out}} / M_{\textrm{in}}$ as the
fraction of the baryonic mass in the wind-affected region
that belongs to bubbles and shells.

Once $F$ is calculated, we determine which individual particles are
``wind'' particles and which ones are not; ``wind'' particles are
ignored when integrating quantities along the LOS and alternative
prescriptions are used to calculate the optical depth contributed by
winds, as we will show in the following. Since there is no exact or
unique way to identify the ``wind'' particles, we decided to flag as
``wind'' a fraction $F$ of all particles in the wind-affected regions,
specifically those with the lowest estimated densities.  This
approximation implies that the particles we remove from the list are
generally residing far from the centre of haloes, while the particles
representing the galaxy and the unperturbed ambient medium tend to
cluster towards the centres. However, since particles have a finite
volume, defined by their SPH smoothing radius $h_{\textrm i}$, the
lowest density particles are also the ones spread over the largest
volumes, and so most likely to contribute to the LOS. Contributions
from matter ``in'' galaxies is mostly unaffected.

In our calculations, we first determine the density and the
density--weighted temperature and velocity contributed by non--wind
particles to each bin $j$ along a LOS (Appendix \ref{optical}). We
then add in the contribution of the winds themselves. Finally, we
calculate the optical depth along the LOS and the flux spectrum. In
Subsections \ref{bb} and \ref{cold} we show how we estimate the
contributions of bubbles and shells respectively and we discuss a few
examples of synthetic spectra.

\subsubsection{Contribution by bubbles}
\label{bb}

\begin{figure}
\centering
\includegraphics[width=8.8cm]{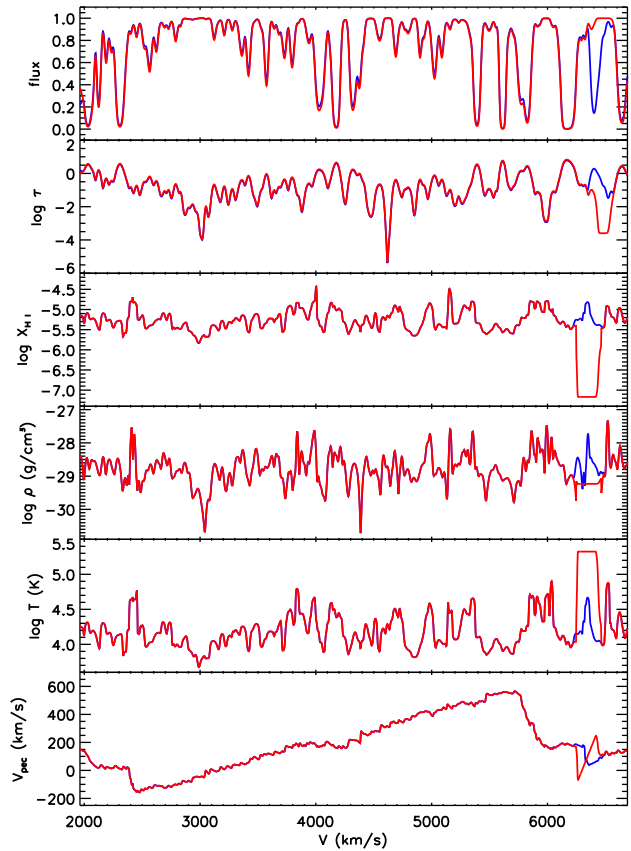}
\caption{Example of spectrum crossing a pressure--driven bubble. The
blue line shows the unperturbed ``no wind'' model, the red line the
e03 wind model. From top to bottom, the different panels reproduce the
flux spectrum, the optical depth $\tau$ of the gas along the line of
sight, the fraction of neutral hydrogen $X_{\hi}$, the total density
$\rho$, the temperature $T$ and the peculiar velocity
$v_{\textrm{pec}}$.}
\label{bubble}
\end{figure}

To calculate the contribution of bubbles to the optical depth along a
LOS, we assume that the material outflowing from the galaxies and the
shocked ambient medium are efficiently mixed and uniformly distributed
inside the bubble. This assumption of a uniform bubble is reasonable
(although not exact), because the sound crossing time of the gas is
shorter than the dynamical time of the wind.

The semi--analytic description of the wind evolution gives information
about the mass, radius, velocity and temperature of the wind. The
temperature of the bubble $T_{\textrm b}$ is determined by the energy
balance in the wind (cfr. eq. 4 of BSW05). Most temperatures
are of order $10^5$ to few times $10^6$ K, which means that the bubble
material is collisionally ionised. We calculate the fraction of
neutral hydrogen $X_{\hi}$ in a bubble by assuming collisional
ionisation equilibrium (Appendix \ref{ion}).

The neutral hydrogen density and the density--weighted temperature and
velocity that a bubble contributes to each bin $j$ which intercepts
the bubble, are:
\begin{equation}\label{densb}
\rho_{\hi} (j) = a^3 X_{\hi} \rho_{\textrm{b}},
\end{equation}
\begin{equation}\label{tempb}
\left(\rho T\right)_{\hi} (j) = a^3 X_{\hi} \rho_{\textrm{b}} T_{\textrm{b}},
\end{equation}
\begin{equation}\label{velb}
\left(\rho v\right)_{\hi} (j) = a^3 X_{\hi} \rho_{\textrm{b}} v_{\textrm{pec,b}},
\end{equation}
where $a$ is the scale factor, $\rho_{\textrm{b}}$ the bubble density and $v_{\textrm{pec}}$ the projection of the peculiar velocity of the wind on the LOS.

In Fig. \ref{bubble} we show an example of the build--up of a spectrum
along a LOS that crosses a bubble. From top to bottom, the panels show
the flux spectrum, the optical depth $\tau$, the fraction of neutral
hydrogen $X_{\hi}$, the total density $\rho$, the temperature $T$ and
the peculiar velocity $v_{\textrm{pec}}$ of the gas along the LOS. The
blue line represents the unperturbed quantities calculated from the
``no wind'' model, while the red line is for the perturbed quantities
from the e03 model.  In this spectrum, the LOS intersects a large
bubble at a small distance from the source galaxy, but does not
intercept the galaxy itself. The ionised gas inside the bubble
produces a decrease of the optical depth, due to its high temperature
and ionisation state. The ``Z'' shape in the velocity distribution of
the gas along the LOS (bottom panel) is due to the outflowing wind
material, $v_{\textrm{s}} \sim 200$ km s$^{-1}$, which modifies the
velocity field around the galaxy.

The flux spectra and the optical depth differ slightly in the two
cases because of normalization effects: this happens when winds modify
the mean optical depth along the LOS by more than a few percent and an
iteration or two is needed during the normalization process. However,
the final difference between the ``wind'' and the ``no wind'' models
after normalization is always less then a few percent, even in the
most extreme cases.

\subsubsection{Contribution by shells and cavities}
\label{cold}

\begin{figure}
\centering
\includegraphics[width=8.8cm]{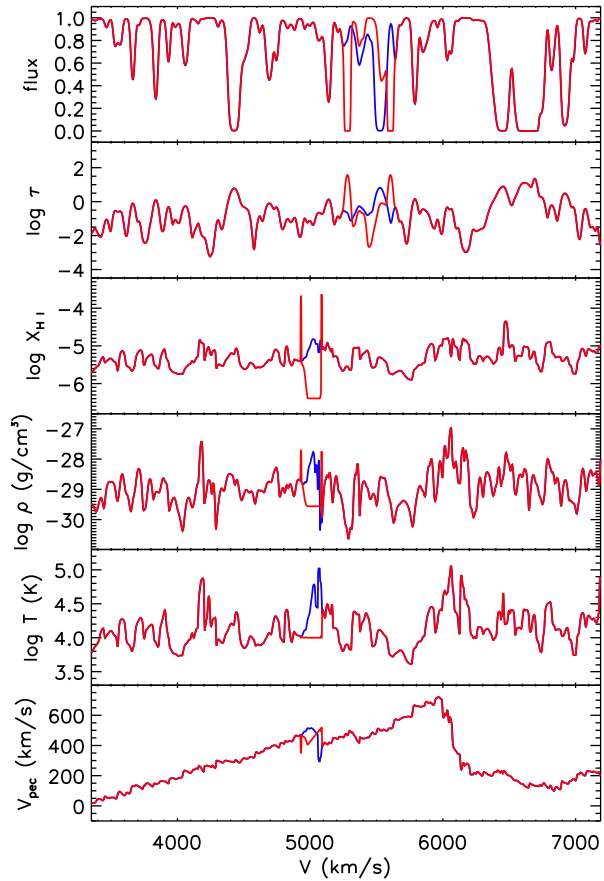}
\caption{Example of spectrum crossing a momentum--driven shell. The
quantities shown are the same as for Fig. \ref{bubble} for the
unperturbed ``no wind'' model (blue line) and for the e03 wind model
(red line).}
\label{shell}
\end{figure}

The effect of shells on the opacity of the gas along a LOS is
substantially different from that of bubbles. Both the
shell and the cavity evacuated in its interior are cold, and have a
more complex density structure. When winds cool down, part of the
outflowing wind mass and all the swept--up mass accumulate in a thin
shell. A cavity is left behind, in which the low--density, cooled wind
material that has not yet reached the shell flows outwards at the wind
velocity. The mass that accumulates in the shell occupies a small
volume and its density can become significantly higher than the
density of the surrounding medium.

We make the assumption that the temperature of shells and cavities is
constant and equal to $T_{\textrm{s}} = 10^4$ K. This assumption is
motivated by the fact that, after the wind cools down and all the
energy is radiated away, the outflowing material reaches
photoionisation equilibrium with the UV background and will maintain
this temperature as long as the density of the shell is not too high.

Our semi--analytic model does not give an explicit description of the
density structure of shells and cavities, but we can reconstruct it by
solving the appropriate set of shock conditions. We use the
Rankine--Hugoniot jump conditions for a non--relativistic shock to
calculate the pressure $P_{\textrm{s}}$ of the gas in the shell. If we
assume that the velocity of the gas behind the shock is equal to the
shock velocity $v_{\textrm{s}}$, we obtain:
\begin{equation}\label{pressure}
P_{\textrm{s}} = P_{\textrm{igm}} + \rho_{\textrm{igm}} \left( v_{\textrm{s}} - v_{\textrm{igm}} \right)^2,
\end{equation}
where $\rho_{\textrm{igm}}$, $v_{\textrm{igm}}$ and $P_{\textrm{igm}}$ are the density, the velocity and the pressure (thermal plus ram) of the ambient medium.
The density of the shell $\rho_{\textrm{s}}$ is then given by the equation of state:
\begin{equation}\label{shelldensity}
\rho_{\textrm{s}} = \frac{\mu m_{\textrm{p}} P_{\textrm{s}}}{k_{\textrm{B}} T_{\textrm{s}}}.
\end{equation}
The radius $R_{\textrm{c}}$ of the cavity internal to the shell is given by:
\begin{equation}\label{cavityradius}
R_{\textrm{c}} ^3 = R^3 - \frac{3 m}{4\pi \rho_{\textrm{s}}}
\end{equation}
and the thickness of the shell is $R_{\textrm{s}} = R -
R_{\textrm{c}}$, with $m$ the shell mass.  The density of the cavity
is determined by the wind mass which has not yet reached the shell
$M_{\textrm{c}}$ and is given by $\rho_{\textrm{c}} = 3 M_{\textrm{c}}
/ 4 \pi R_{\textrm{c}} ^3$.

We estimate the contribution of shells and cavities to the neutral
hydrogen density and the density--weighted temperature and velocity
along the LOS by using eqs. (\ref{densb}), (\ref{tempb}) and
(\ref{velb}), where the index ``$b$'' has been changed with ``$c$''
and ``$s$''.  Shells affect only a few bins on the outer edge of a
wind penetrated by a LOS. The number of the bins is determined by the
thickness of the shell, which is normally as large as the linear
dimension of few bins (typically, the bin width is about 20--26
$h^{-1}$ kpc). The remaining bins inside the shell get contributions
from the cavity gas.

In Fig. \ref{shell} we give an example of the build--up of a spectrum
that crosses a shell. The quantities shown are the same as in
Fig. \ref{bubble}. The shell crossed by the LOS has a density about
ten times the density of the ambient medium, but the fraction of \hi\
in the shell is about 100 times higher. The cold neutral gas in the
shell produces two almost symmetric spikes in the optical depth, which
appear as two narrow and saturated absorption lines in the flux
spectrum. The temperature of $10^4$ K that we assume for the shell
determines the small width of the lines, while the high neutral
fraction produces the saturation.

The LOS in Fig. \ref{bubble} and \ref{shell} show examples of the
effects of winds in spectra. In general, a line of sight may not cross
any wind at all or it may cross several, in which case the spectrum
may become more confused.

\section{Flux PDF}
\label{fpdf}

\begin{figure}
\centering
\includegraphics[width=8.8cm]{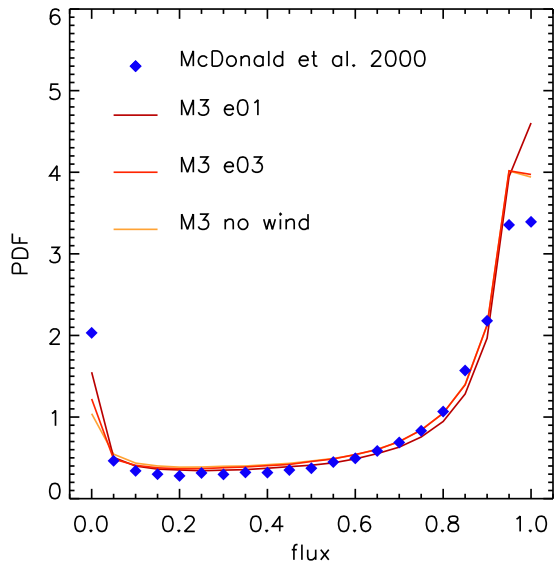}
\caption{Flux PDF for the \lya\ forest at $z=3$. The numerical results
are compared to the observations of \citet{mcd}.}
\label{pdf}
\end{figure}

We calculate the flux probability distribution function (PDF) as a
first comparison between our simulated spectra and the
observations. We use about 600 spectra from our random sample at $z=
3$ and we compare the results with the observations of
\citet{mcd}. Fig. \ref{pdf} shows our results for the ``no wind'' model
and for the e01 and e03 wind models.  The agreement between the
observations and our three sets of simulations is in general very good
and deviations from the observed values are significant only at fluxes
$F \sim 0$ and $F > 0.9$.
However, the observed values corresponding to $F \geqslant 0.9$ are not fully reliable, because they can be strongly affected by continuum fitting errors. 

In general, the e03 model gives almost exactly the same result as the
``no wind'' model, while a small but more significant difference is
evident with the e01 model. This may be an indication that the weak
winds predicted by the e03 model are unable to affect the flux PDF of
the \lya\ forest, while more efficient winds as in the e01 model may
slightly modify it.  However, in all cases the difference between the
numerical results and the observations are tiny, and not sufficient to
rule out any of the feedback models.

\section{Power spectrum}
\label{fpower}

\begin{figure}
\centering \includegraphics[width=8.8cm]{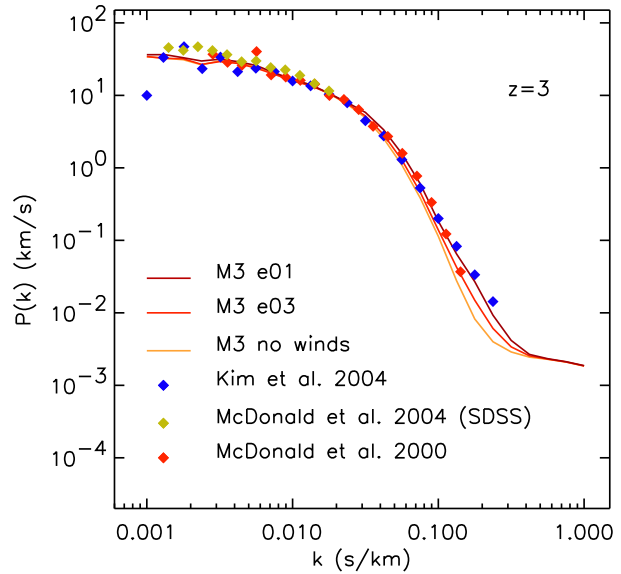}
\caption{The flux power spectrum computed from synthetic spectra at
$z\sim 3$. The numerical results are compared to the observations of
\citet{kim} at $z\sim 2.54$, \citet{mcd} at $z=3$ and \citet{mcsdss} at
$z=3$. The synthetic spectra have been calculated using the samples of
randomly selected LOS for the ``no wind'', e01 and e03 models.  }
\label{kim}
\end{figure}

To calculate the power spectrum of the flux, we use the estimator
proposed by \citet{hui}.
Since the power spectrum is sensitive to
changes in the mean spectral flux $\langle e^{-\tau} \rangle$, which
is a function of redshift, \citet{hui} suggest calculating the power
spectrum of the quantity:
\begin{equation}
F = \frac{e^{-\tau}}{\langle e^{-\tau} \rangle} -1,
\end{equation}
where $\tau$ is the optical depth and $e^{-\tau}$ the amplitude of the
fluctuations.  Fig. \ref{kim} shows our results for the power spectrum
of the transmitted flux $P(k)$, calculated in this way from the sets
of random spectra at $z=3$ for the ``no wind'', e01 and e03
models.
All the spectra for the three wind models have been calculated assuming the same relationship between the gas density and temperature, as given in Subsection \ref{gas}. This is important, because the flux spectrum is sensitive to changes in the density--temperature relationship: the same gas conditions ensure that our results show variations due to winds and not to simulation artefacts.

The numerical results are compared to the observations of
\citet{kim}, \citet{mcd} and \citet{mcsdss}.  \citet{kim} and
\citet{mcd} compute the power spectrum from few high resolution quasar
spectra taken with the VLT and Keck telescopes respectively, while
\citet{mcsdss} use a sample of about 3000 medium resolution spectra
from the Sloan Digital Sky Survey (SDSS).  The lower resolution of the
SDSS spectra limits the estimate of $P(k)$ in the range $0.0013-0.04$
s km$^{-1}$, while higher resolution spectra allow to estimate $P(k)$
in the broader range $0.001-0.2$ s km$^{-1}$. At $k>0.3 - 0.4$ s
km$^{-1}$ $P(k)$ is dominated by noise and flattens out.

In Fig. \ref{kim} there is agreement between the ``no wind''
model and the e01 and e03 wind models at $k<0.1$ s km$^{-1}$,
with deviations appearing only at smaller scales. At $k\gtrsim 0.1$ s
km$^{-1}$ the three numerical models diverge and, broadly speaking,
the higher the wind efficiency, the more power is contributed at
smaller scales. This is consistent with our previous results for the
wind filling factor and the fraction of IGM mass in winds
(BSW05), which predict that, although winds may not heavily
modify the statistical properties of the \lya\ forest, the largest
deviations should be found in the e01 model.  In addition, the fact
that more efficient winds contribute more power on small scales is
consistent with the presence of the narrow and saturated absorption
lines created by shells, which are more numerous in the e01
model. These lines contribute power to scales smaller than about a
hundred kiloparsecs.

Unfortunately, the spectral region corresponding to $k\gtrsim 0.1$ s
km$^{-1}$ is also the region where unidentified metal lines
(principally $\textrm{O \small{VI}}$ and $\textrm{C \small{IV}}$) blended with the \lya\ forest might give a substantial contribution to the power.
\citet{kim} estimate that between 20 and 70\% of the power is contributed by metal lines for $k\gtrsim 0.1$ s km$^{-1}$.
Although we are tempted to
suggest that both our wind models seem to fit the data better than the
``no wind'' model, we cannot definitely rule out the ``no wind'' model
itself, since our current \lya\ spectra do not include metal lines and
we are unable to estimate well how much power is contributed by them.

The good agreement between the two wind models and the ``no wind''
model at $k <0.1$ s km$^{-1}$ is an indication that the effects of
galaxy feedback on the \lya\ forest, and in particular of galactic
winds, should not affect the estimation of cosmogical parameters.

\section{Flux transmissivity around LBG}
\label{fadel}

\begin{figure}
\centering
\includegraphics[width=8.8cm]{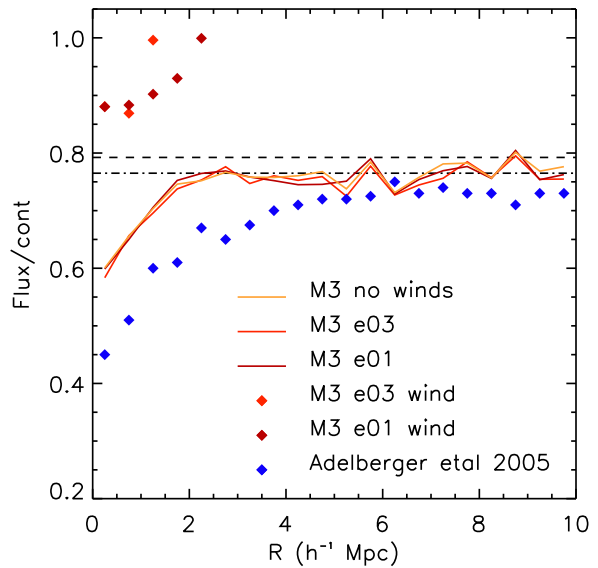}
\caption{The mean flux transmissivity around LBG as a function of the
impact parameter. Results are presented for a sample of about 7700
simulated galaxies with random impact parameters (lines) and for
subsamples of galaxies blowing winds with radii larger than the impact
paramater (diamonds). The observational data are from
A05.}
\label{adel}
\end{figure}

\begin{figure}
\centering
\includegraphics[width=8.8cm]{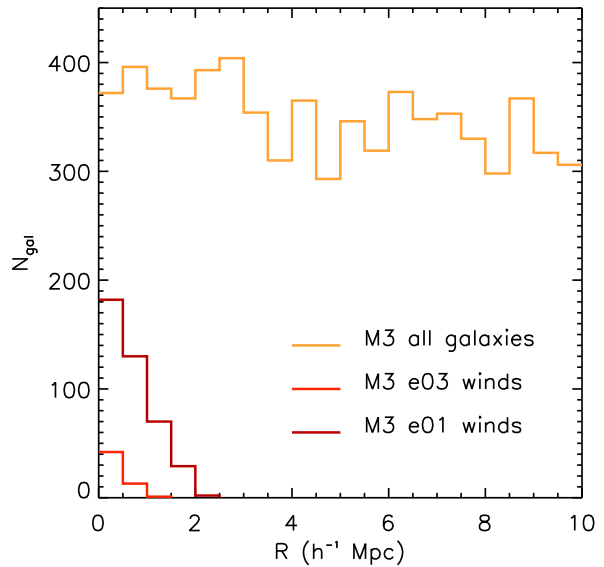}
\caption{The total number of galaxies and the number of wind--blowing
galaxies with wind radii larger than the impact parameter as a
function of impact parameter in the simulated sample.}
\label{histo}
\end{figure}

\begin{figure*}
\centering
\includegraphics[width=0.49\textwidth]{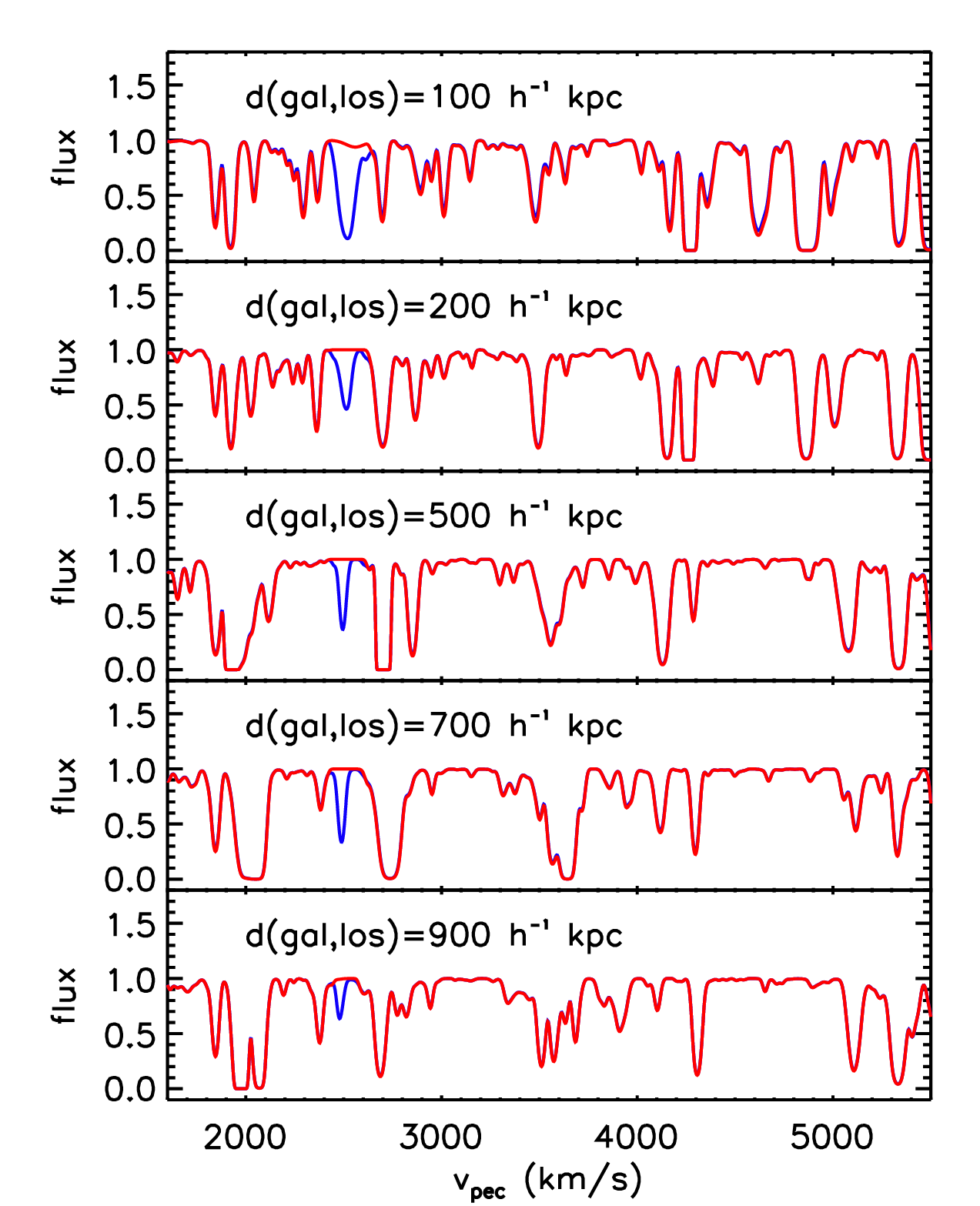}
\includegraphics[width=0.49\textwidth]{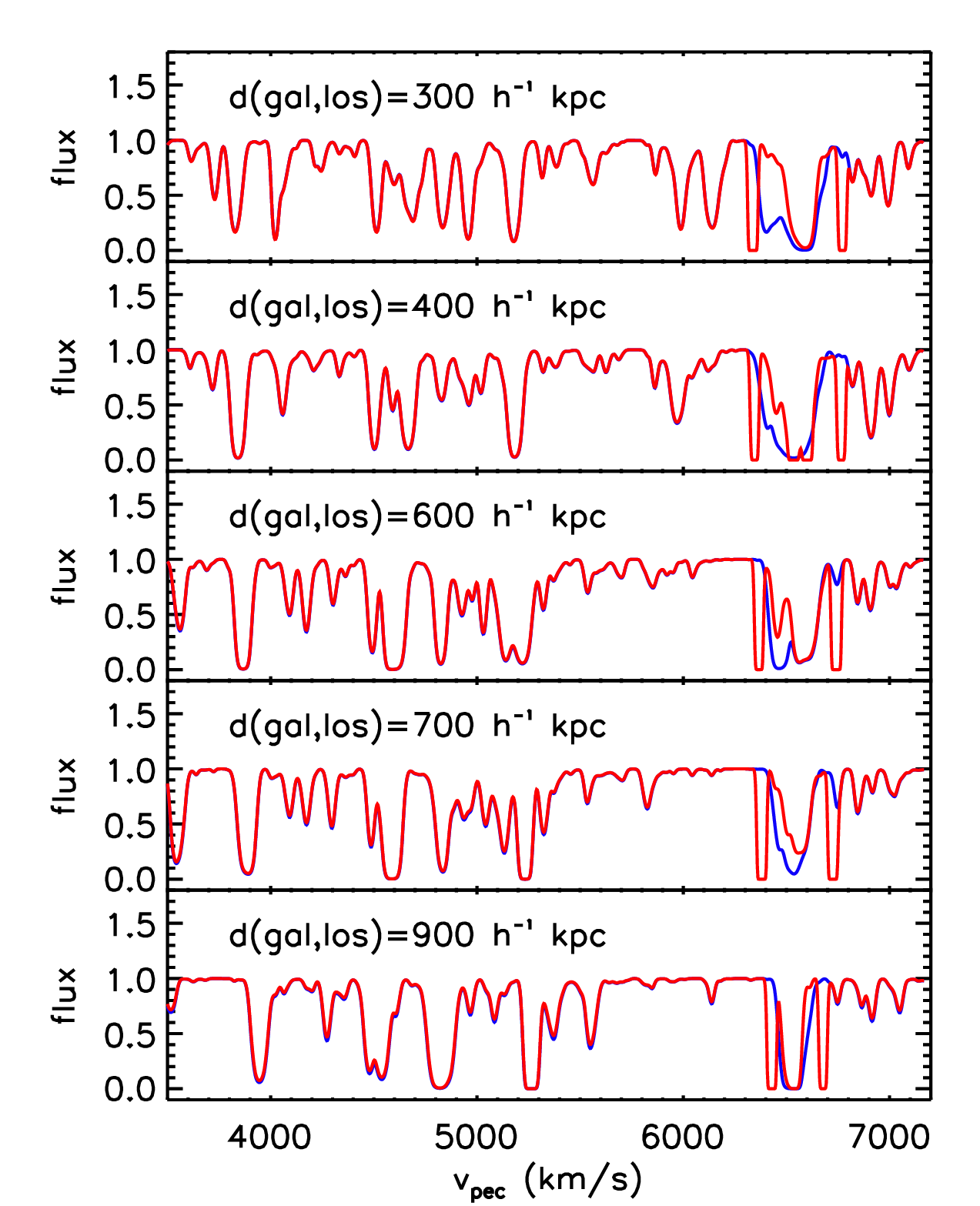}
\caption{Two sets of parallel lines of sight intercepting isolated
field galaxies blowing winds into the IGM at impact paramaters $100 <
b < 900$ $h^{-1}$ comoving kpc. The left panel shows how the
absorption from neutral gas surrounding a galaxy is quenched inside a
hot pressure--driven bubble. The right panel shows the effect of a
momentum--driven wind and the absorption by the cold gas in the
shell. The red line represents the spectrum perturbed by winds from
the e03 model, while the blue line the unperturbed spectrum from the
``no wind'' model. The galaxy in the left panel has
$M_{\star}=1.6\cdot 10^9$ $h^{-1}$ M$_{\sun}$ and is blowing a bubble
with radius $R = 257$ $h^{-1}$ (physical) kpc, velocity 274 km
s$^{-1}$ and temperature $T=2.2\cdot 10^5$ K. The galaxy in the right
panel has $M_{\star}=9.2\cdot 10^9$ $h^{-1}$ M$_{\sun}$ and is pushing
a shell with radius $R = 272$ $h^{-1}$ (physical) kpc and velocity 103
km s$^{-1}$.}
\label{d2}
\end{figure*}

In this Section we present our results for the mean \lya\ flux
transmissivity near Lyman break galaxies and we investigate the effect
of winds on the absorption spectra at small impact parameter. We use
our sample of galaxy--selected spectra at $z=3$ and $z=2$ for the ``no
wind'', e01 and e03 models (Section \ref{loes}).

Fig. \ref{adel} shows our results for the mean transmitted flux as a
function of the impact parameter $R$. The numerical results are
compared to the observations of A05. The analysis of our
full set of spectra confirms the observation that the mean \lya\
transmissivity near LBGs decreases for small impact parameters.  Our
spectra yield a smaller decrease than seen in the observations: this
is partly due to the fact that our semi--analytic prescriptions are
unable to describe the cold neutral gas in haloes and simply assume
that this gas is at the virial temperature and so highly ionised.  In
addition, the mean flux $\langle F \rangle = 0.792$ in the simulated
spectra is higher than in the data, where $\langle F \rangle = 0.765$.
This is not a severe problem, however, because the difference is small
and all our models do predict a non--negligible decrease in the \lya\
flux.

A more detailed analysis shows that the \lya\ flux is almost
completely transmitted around a significant fraction of the galaxies
in our model and that the gas in these regions has an extremely low
optical depth. This is shown in Fig. \ref{adel} by the diamonds in the
upper left corner, which represent the mean transmitted flux along
those LOS that actually intercept wind bubbles.  These galaxies show
transmitted fluxes up to 50--60\% higher than the mean flux found for
the full sample.

In Fig. \ref{histo} we show the total number of LOS and the number of
intercepted wind bubbles in the e01 and e03 models as a function of
impact parameter. Since we assign random impact parameters when
extracting the LOS, the distribution of the total number of galaxies
is close to uniform. The e01 model produces wind--blowing galaxies
with bubble radii as large as 2.5 $h^{-1}$ Mpc, while fewer winds and
maximum radii of 1.5 $h^{-1}$ Mpc are produced by the e03 model. The
fraction of galaxies for which the LOS intercepts a bubble for impact
parameters below 1 $h^{-1}$ Mpc is about 7\% for the e03 model and
about 40\% for the e01 model. These numbers are in good agreement with
the finding of A05 that about a third of the galaxies in
their sample show almost complete transmission of the quasar \lya\
flux at the LBG redshift.

Winds either evacuate cavities or drive bubbles of hot gas into their
surroundings. Although the physics of pressure--driven and
momentum--driven winds is different, in either case the optical depth
of the gas may be reduced inside the wind, and the flux increased. In
the case of adiabatic winds, the hot gas in bubbles is highly ionised,
bubble temperatures being normally higher than $10^5$ K. In case of
momentum--driven winds, the swept--up gas accumulates into a thin
shell, while a cavity is evacuated in the interior, where only the
cooled wind is flowing outwards. The density of the residual gas
inside the cavity is tiny and its contribution to the optical depth
along the LOS is generally negligible. Occasionally another absorber
external to the galaxy may intervene to produce an uncorrelated
absorption line.

In Fig. \ref{d2} we show the effects of winds on artificial spectra at
various impact parameters from source galaxies in the e03 wind
model. For clarity, we have chosen two isolated field galaxies with
wind radii larger than 1 $h^{-1}$ comoving Mpc and with no other
wind--blowing galaxies in their proximity. In regions where galaxies
are more clustered, winds overlap and their signatures in spectra mix
and become less easily identifiable.  The left panel shows the effect
of a pressure--driven bubble of hot gas that ionises the gas
surrounding the source galaxy (see also Fig. \ref{bubble}). Since
there are no intervening absorbers along the LOS, the bubble increases
the flux transmissivity in the spectral region corresponding to the
galaxy position.  The right panel shows a momentum--driven shell that
evacuates a cavity and accumulates material at its edges (see also
Fig. \ref{shell}).  The symmetric pattern of two narrow and saturated
absorption lines is clearly visible. The separation between the two
absorption lines and the pixel corresponding to the minimum distance
to the galaxy is roughly equal to the velocity of the shell projected
onto the LOS.

\begin{figure}
\centering
\includegraphics[width=8.8cm]{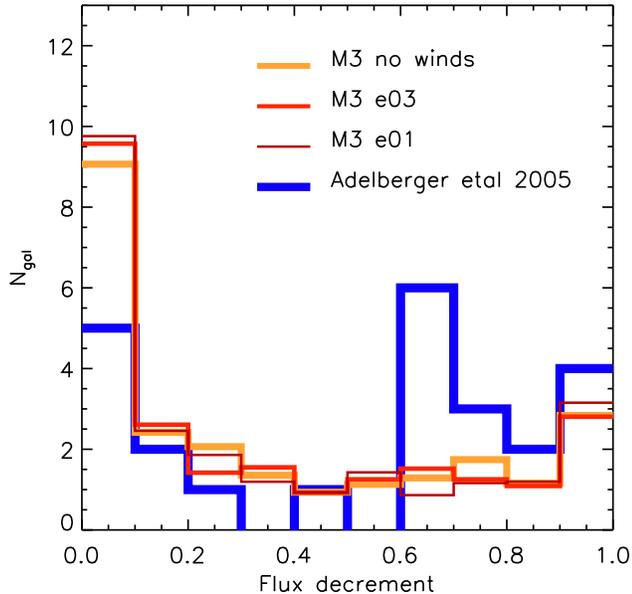}
\caption{The distribution of galaxies with impact parameter $b<1$ $h^{-1}$ Mpc
as a function of the flux decrement observed in the spectra. The
observational data are from A05.}
\label{mega}
\end{figure}

In Fig. \ref{mega} we show the distribution of galaxies with impact
parameter smaller than 1 $h^{-1}$ Mpc as a function of the flux
decrement.  The numerical results are compared with the data of
A05 and normalized to the total number of objects in the
observational sample (e.g. 24 galaxies). The flux decrement is defined
as $F_{\textrm{D}} = 1-F$.

A05 claim that numerical simulations are unable to reproduce the high
number of galaxies with $F_{\textrm{D}} \sim 0$ found in their
data. In Fig. \ref{mega} we show that this is not the case for our
model, independent of the wind parameters. In fact, the
distribution of galaxies as a function of the flux decrement is an
alternative way to represent the PDF of the flux in the spectra and,
if the impact parameters of the LOS are randomly chosen in the
interval $(0, 1)h^{-1}$ Mpc, then the distribution of galaxies in
Fig. \ref{mega} should qualitatively look like the PDF of the flux in
Fig. \ref{pdf}, where the $x$--axis has been inverted from $F$ to
$F_{\textrm{D}} = 1-F$.  The simulation result presented by
A05 that almost no galaxies have $F_{\textrm{D}} \sim 0$
could be understood if the impact parameter of the galaxies in the
sample is close to 0 or smaller than the virial radius. In this case,
a LOS would intercept the reservoir of neutral hydrogen in the
halo/disk of the galaxy and a large flux decrement should be expected.
Although our galaxy distribution explains the relatively high number
of galaxies with $F_{\textrm{D}} \sim 0$, a more detailed comparison
between our data and the observations is inconclusive. In
fact, the 24 observed galaxies are too few to define a distribution
calculated over 10 bins.  For comparison, we have almost 800
galaxies in our model sample.  The e01 and e03 wind models predict a
slightly larger number of galaxies with small flux decrements than the ``no
wind'' model, in agreement with what we showed in
Fig. \ref{adel}.

\section{Gap statistics}
\label{gap}

\begin{figure*}
\centering
\includegraphics[width=0.49\textwidth]{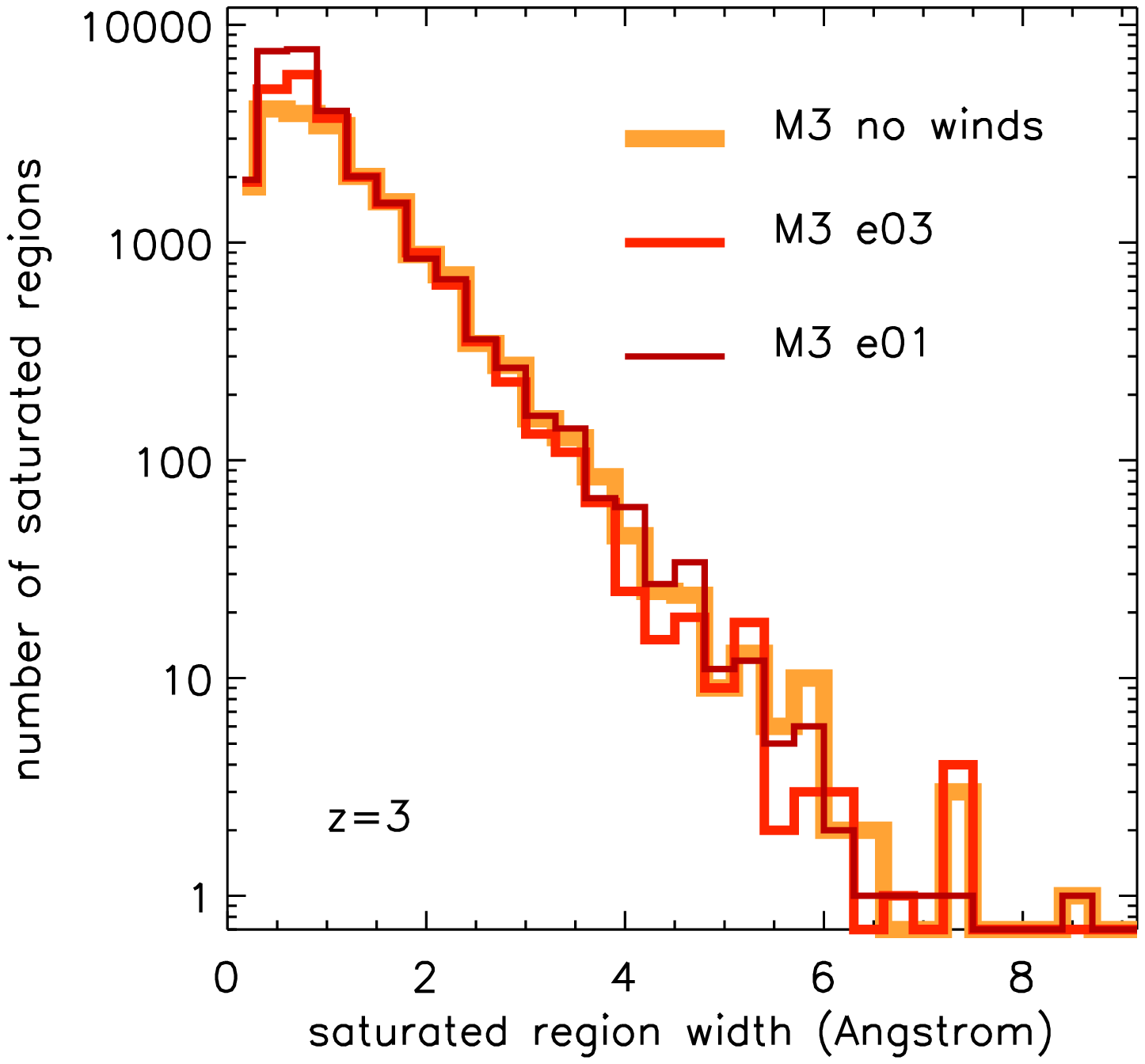}
\includegraphics[width=0.49\textwidth]{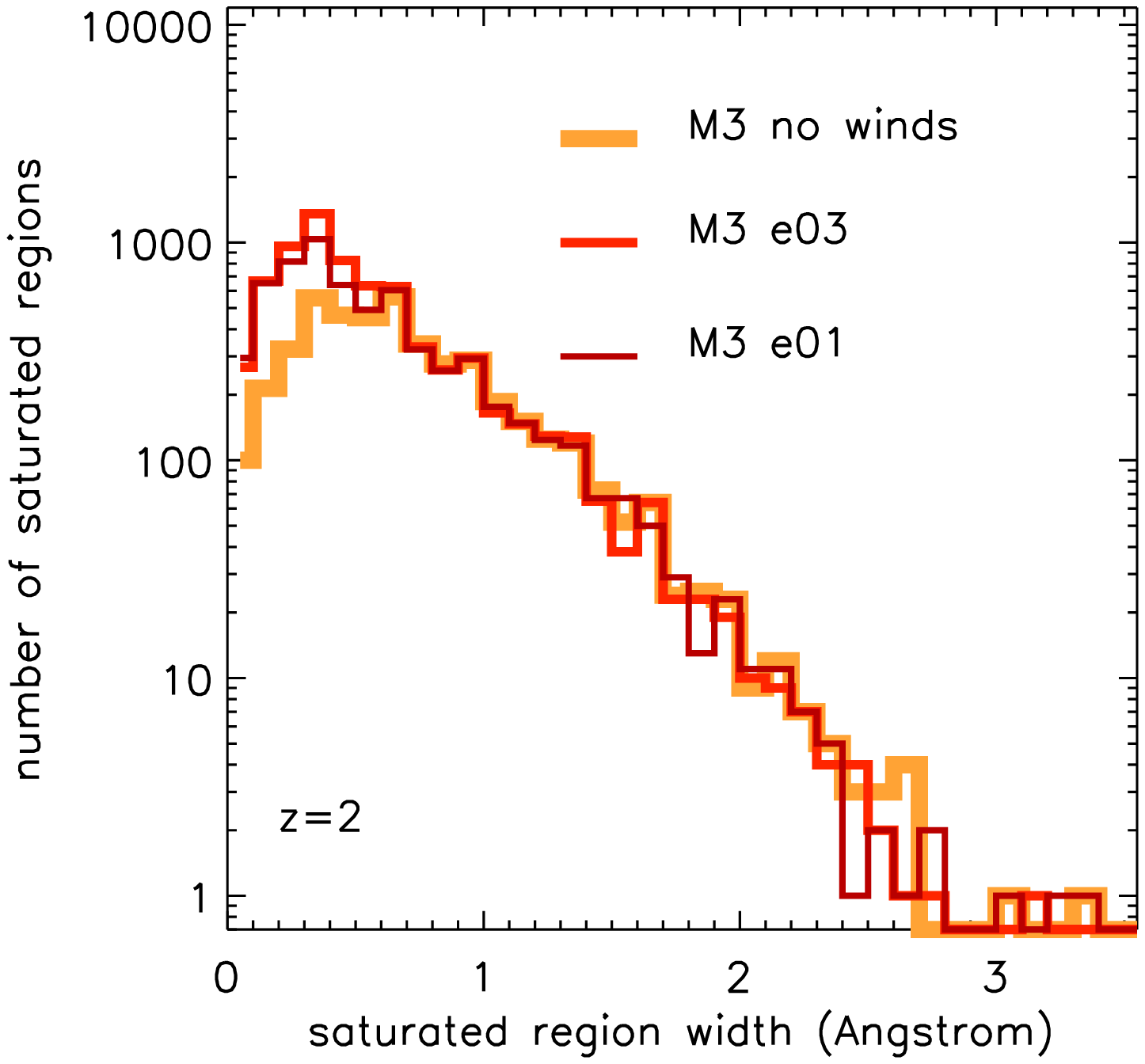}
\caption{The number distribution of saturated regions as a function of
region width. Results are shown at $z=3$ (left panel) and $z=2$ (right
panel). Shells contribute a large number of saturated lines with
widths smaller than about 1 \AA .}
\label{void}
\end{figure*}

\begin{figure*}
\centering
\includegraphics[width=0.49\textwidth]{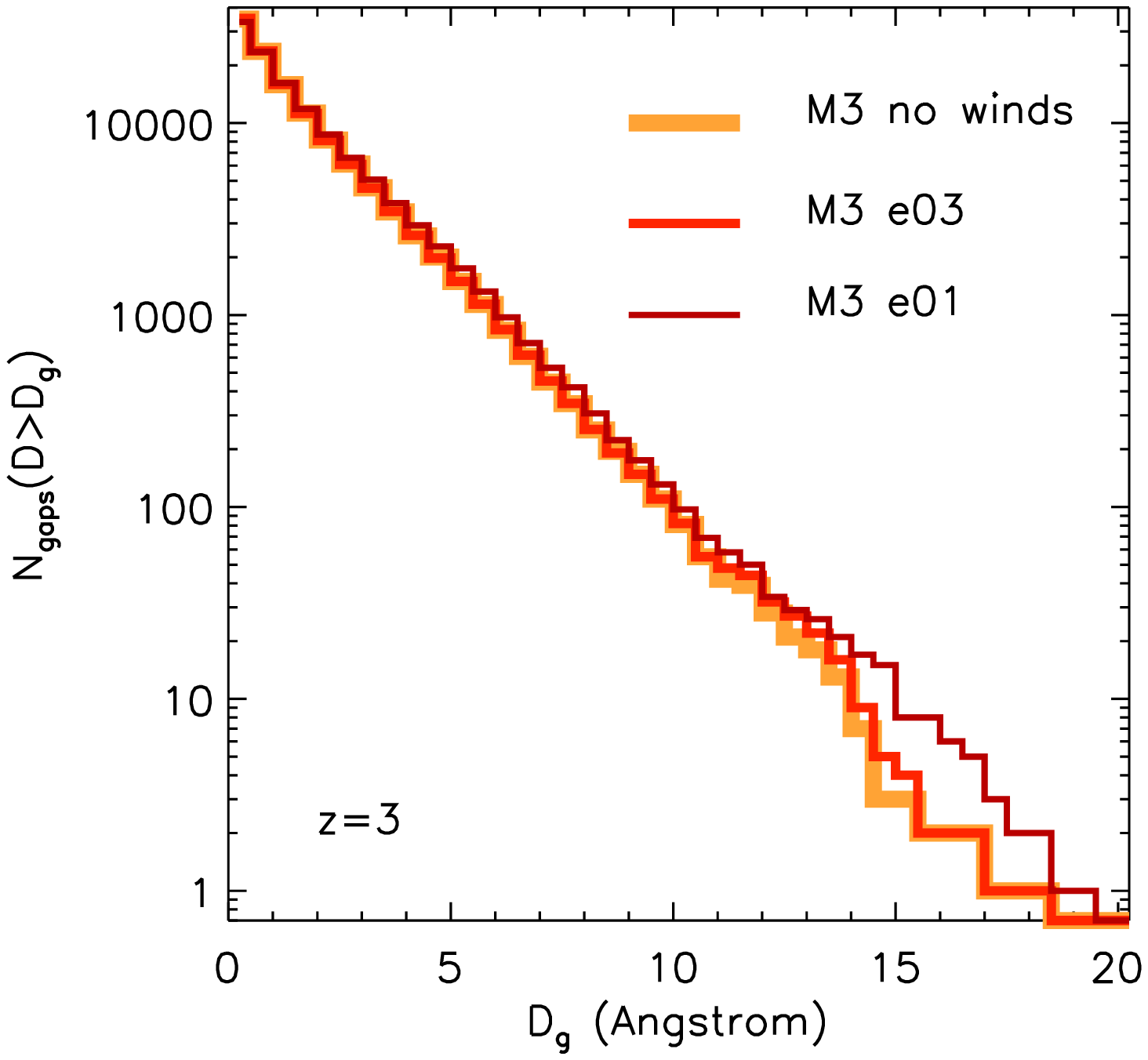}
\includegraphics[width=0.49\textwidth]{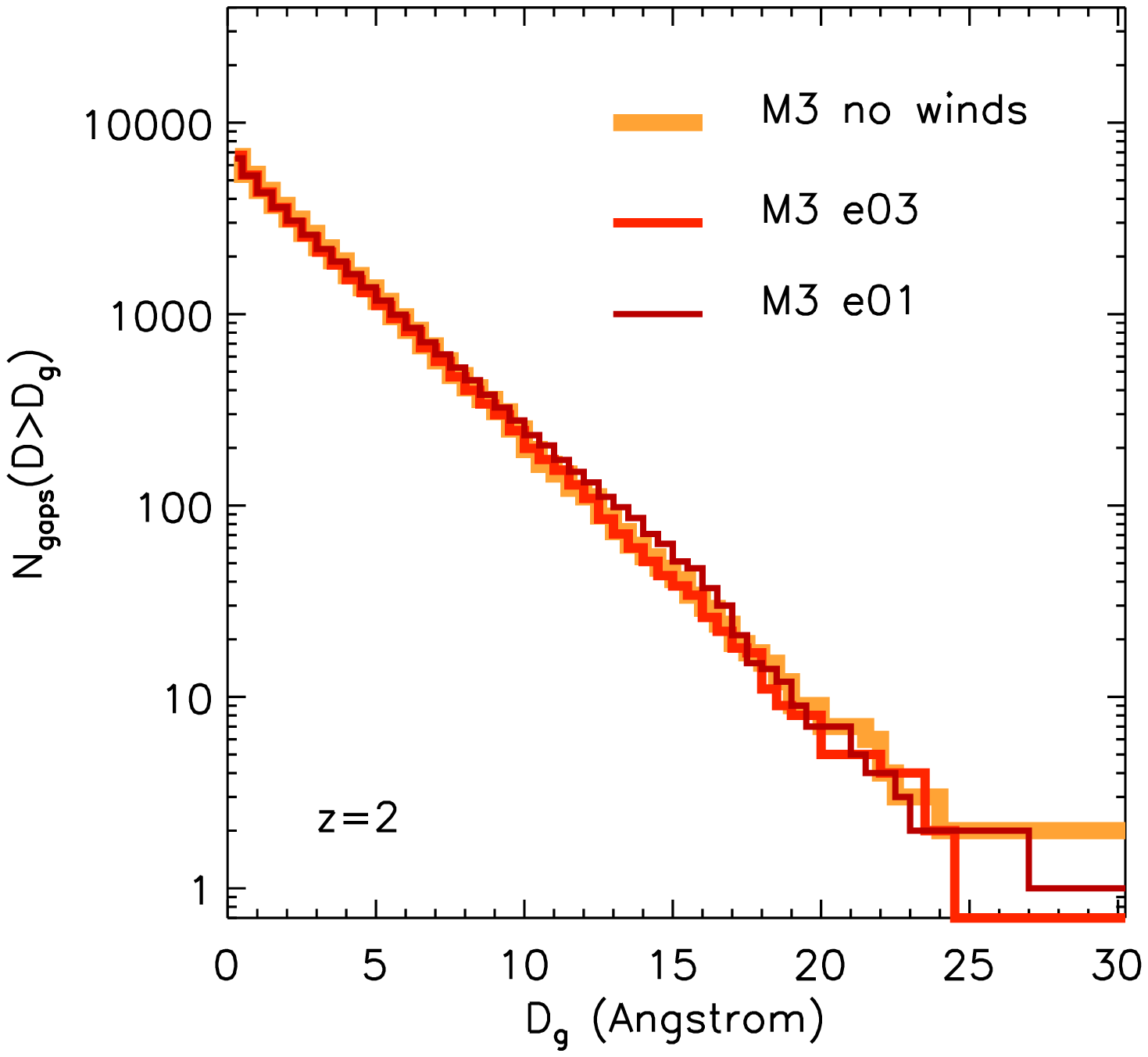}
\caption{The cumulative distribution of gaps in spectra as a function of
gap width. Results are shown at $z=3$ (left panel) and $z=2$ (right
panel). The distribution of gaps in the e01 model is mostly affected for gap widths $10 \lesssim D_{\textrm g} \lesssim 20$ \AA\, while the e03 model shows a much smaller deviation from the ``no wind'' model.}
\label{gapgap}
\end{figure*}

We showed in the previous sections that winds do leave specific
signatures in the \lya\ forest, although they may not strongly affect
the flux PDF and the flux power spectrum. In this section, we present
a new method to test statistically for the presence of wind absorption
lines in the \lya\ forest. The method is based on two findings: first,
when a LOS crosses a shell, a narrow and saturated absorption line is
created; secondly, these absorption lines due to shells appear to have
roughly all the same line width and steep profile.  We look for shell
absorption lines in our artifical spectra by identifying all the
spectral regions where absorption is complete. To select such regions,
we require that the flux is smaller than $F < 0.1$ throughout. We
estimate the width in wavelength of each region, which we call
$D_{\textrm s}$, and finally we calculate the distribution of the
saturated regions as a function of the region width $D_{\textrm s}$.

Our results are shown in Fig. \ref{void} for $z=3$ and $z=2$.  The
distribution of saturated regions is not affected by winds except for
very small values of $D_{\textrm s}$, where absorption lines due to
shells in the wind models almost double the number of saturated
regions with $0.3 < D_{\textrm s} < 1$ \AA\ at $z=3$.  At $z=2$, where
the number and the width of saturated regions shrinks significantly
because of the cosmic expansion and the increased intensity of the UV
background radiation, the wind models produce almost three times as
many lines with $0.1 < D_{\textrm s} < 0.6$ \AA\ as the ``no wind''
model.  This analysis cannot be directly translated into a
distribution of lines as a function of column density, because the
width of saturated regions $D_{\textrm s}$ is by no means a direct
estimate of the column density of the absorber. However, our results
show that the presence of absorption lines due to shells should
significantly increase the number of narrow saturated absorption lines.

We then tried a similar analysis to test for the presence of
bubbles. In analogy to what we did for shells, we start from the
result that bubbles may enhance the transmissivity of the \lya\ flux
in their interior, sometimes completely washing out the absorption by
neutral gas. We therefore look for ``gaps'' in the \lya\ forest and we
identify all the regions with fluxes $F > 0.9$, calling their width
$D_{\textrm g}$. We then calculate the cumulative distribution of the gaps as a
function of this width.  If bubbles had a large effect on the
distribution, we would expect to find an increase in the number of
gaps at the largest widths. The effect should be stronger at $z=3$
than at $z=2$, since the gap width in the absence of bubbles tends to
be larger at lower redshift because of the lower mean opacity of the IGM.

As can be seen in Fig. \ref{gapgap}, we do find such an effect, which is particularly significant for the e01 model. This model predicts up to 50\% more gaps with $10 < D_{\textrm g} < 20$ \AA\ and about 5--20\% more gaps with $5<D_{\textrm g} <10$ \AA\ at both redshifts. For $10 < D_{\textrm g} < 20$ \AA\ the e03 model shows a maximum deviation of about 20\% from the ``no wind'' model at $z=3$ and almost no deviation at $z=2$. At smaller gap widths there is no difference from the ``no wind'' model.
The larger deviations from the ``no wind'' model in the results for the e01 model can be understood when two main properties of the winds are considered: i) the total number of wind--blowing galaxies is about ten times larger in the e01 model than in the e03 model; ii) the fraction of pressure--driven bubbles in the e01 model is two times larger than in the e03 model. The maximum radius of bubbles in the e01 model is also larger than in the other model.

The effect of bubbles may be difficult to distinguish in real spectra. The unperturbed cumulative distribution of gaps is unknown and no direct comparison is available. Metal lines may somewhat alter the distribution itself, but this problem could in principle be overcome, because only high column density metal lines not blended in the \lya\ forest would affect the distribution, and these should be relatively straightforward to identify and remove.
Althought the first problem is more difficult to address, we believe that the effect of bubbles in spectra could still be successfully investigated. In fact, in Fig. \ref{gapgap} the cumulative distribution of gaps in unperturbed spectra closely follows a power law relation, while wind models introduce a ``bump'' in the distribution for $10 < D_{\textrm g} < 20$ \AA . This deviation from a power law, as we mentioned in the previous paragraph, can be as high as 50\% for the e01 model and it could be used to test for the presence of winds in spectra: on the basis of our result, if the cumulative distribution follows a power law, most likely there are no bubble signatures in spectra; on the other hand, if a bump does appear in the cumulative distribution at $10 < D_{\textrm g} < 20$ \AA , then hot winds may leave a detectable footprint in the forest.

\section{Conclusions}
\label{conclu}

In this paper, we have used numerical simulations to investigate the
effect of galactic winds on the \lya\ forest. 
Our main conclusions can be summarized as follows:

\begin{enumerate}

\item The flux PDF calculated from simulated spectra agrees well with
the flux PDF of \citet{mcd} at $z=3$ for all three wind models. The
effects of the winds are barely discernible the PDF. The small
deviations observed between the numerical results and the
observational data are insufficient to rule out any of the models.

\item Our simulated power spectra of the \lya\ flux at $z=3$ agree
well with the observational results of \citet{kim}, \citet{mcd} and
\citet{mcsdss} on all scales, with the e01 wind model closely matching
the \citet{kim} data and the e03 model best fitting the data of
\citet{mcd} on small scales. The ``no wind'' model tends to
underestimate the power on the smallest scales.  Winds sensibly
increase the power on scales $k\gtrsim 0.1$ s km$^{-1}$ at $z=3$, when
both the volume filling factor $f_{\textrm v}$ and the fraction of IGM
mass in winds $f_{\textrm m}$ are small. The overall effect is not
large, however, and deviations affect only the spectral region which
is already contaminated by metal lines.  Since $P(k)$ is not affected
by winds on large scales ($k<0.1$ s km$^{-1}$), the estimation of
cosmological parameters using the \lya\ forest flux power spectrum
should be insensitive to galactic winds.

\item The \lya\ flux transmissivity is enhanced near wind--blowing
galaxies, independent of whether the wind is creating a bubble of hot
gas or a cooled cavity surrounded by a shell. This is consistent with
the observations of A05, who find that up to a third of
all the galaxies in their sample with impact parameter smaller than $1
h^{-1}$ Mpc have transmitted fluxes up to 90\%. The mean \lya\
transmissivity decreases with decreasing impact parameter, in
agreement with previous numerical models and the A05
data.

\item To search for wind signatures in quasar spectra, we have devised
a new method which measures the abundance of saturated regions as a
function of their width. We find that absorption due to wind shells
increases the expected number of saturated regions with $D_{\textrm s}
\lesssim 1$ \AA\ by a factor of two or more at $z=3$ or 2, forcing the
distribution to behave as a power law down to smaller widths than in
the ``no wind'' model. In fact, the ``no wind'' model shows a
turn--over at $D_{\textrm s} \sim 1$ \AA\ at $z=3$ and at $D_{\textrm
s} \sim 0.6$ \AA\ at $z=2$, while the wind models have such a point at
$D_{\textrm s} \sim 0.5$ and $D_{\textrm s} \sim 0.2$ \AA\
respectively. Although not a huge effect, this is certainly large
enough to be detectable. Since we consider saturated regions widths
rather than line column densities, this measure has the advantage of
being insensitive to unsaturated metal lines blended with the \lya\
forest.

\item A similar analysis aimed to estimating the abundance of gaps in
the absorption pattern of the \lya\ forest shows a weaker but still detectable effect from wind bubbles. We have calculated the cumulative distribution of gaps as a function of the gap width and we find that winds contribute as much as 50\% more gaps with $10 < D_{\textrm g} < 20$ \AA\ in the e01 model. We suggest that this effect may be used to test the presence of hot winds in the \lya\ forest: if the cumulative distribution of gaps closely follows a power law, then it is likely that no winds are present, while if a bump appears for $10 < D_{\textrm g} < 20$ \AA , then, according to our model, the \lya\ forest does contain wind signatures.

\item The two new methods we present in Section \ref{gap} do not tell us directly \emph{where} wind signatures are, but only \emph{whether} they are present or not. However, this is already a significant improvement with respect to previous works.
Finally, the detection of a large bump in the mentioned range may give us an indication that the powerful winds with low entrainment fractions of the e01 model are more likely to occur in the Universe than the weaker and more mass loaded winds in the e03 model.
\end{enumerate}

\section*{Acknowledgments}
We would like to thank M. Viel and T. Theuns for help in constructing and analysing the spectra and M. Haehnelt, T.--S. Kim, F. Stoehr and P.A. Thomas for interesting discussions and suggestions that improved the manuscript.
We also thank the anonimous Referee for useful suggestions.
This work has been supported by the Research and Training Network ``The Physics
of the Intergalactic Medium'' set up by the European Community under contract
HPRN--CT--2000--00126.
SB was partially supported by PPARC.

\appendix

\section{Hydrogen ionisation state}
\label{ion}

To calculate how much radiation of a desired wavelenght is absorbed by
a gas cloud, we need to know the ionisation state of the gas, which
determine its optical depth.  To calculate the fraction of neutral
hydrogen $X_{\hi}$ in a gas, we solve the equation for ionisation
evolution of hydrogen. In the limit of ionisation equilibrium and for
a highly ionised gas, we find (\citealt{cen}, \citealt{tom}):
\begin{equation}\label{ionisstate}
X_{\hi} = \frac{\alpha_{\hii} n_{\textrm e}} {\Gamma_{\gamma \hi} + \Gamma_{{\textrm e} \hi} n_{\textrm e}},
\end{equation}
where $n_{\textrm e}$ is the electron number density, $\alpha_{\hii}$
the recombination rate of hydrogen, $\Gamma_{\gamma \hi}$ the
photoionisation rate of hydrogen and $\Gamma_{{\textrm e} \hi}$ the
collisional ionisation rate.  The recombination rate of hydrogen is a
function of the temperature of the gas and, defining $T_{\textrm n} =
T/(10^n \textrm{ K})$, is given by  \citep{haardt}:
\begin{equation}
\alpha_{\hii} = \left\{
  \begin{array}{ll}
  1.58\cdot 10^{-13} T_{4} ^{-0.51}, &
  T \leq 10^4 \textrm{ K} \\
  1.58\cdot 10^{-13} T_{4} ^{-0.51 - 0.1 \log T_{4}}, &
  T > 10^4 \textrm{ K} \\
  \end{array} \right.
\end{equation}

Hydrogen is ionised through two different mechanisms:
collisional ionisation and photoionisation. Photoionisation is the dominant
process in the low density gas with $T \lesssim 10^5$ K.
The photoionisation rate depends on the flux spectrum of the ionising UV background radiation $J(\nu, z)$:
\begin{equation}
\Gamma_{\gamma \hi} (z) = \int^{\infty} _{\nu _{\h}} \frac{4\pi J(\nu ,z) \sigma_{\h} (\nu)}{h\nu} \di \nu ,
\end{equation}
where $\sigma_{\h} (\nu)$ is the photoionisation cross--section and $\nu_{\h}$ the ionising threshold frequency of hydrogen.
We use the ionising UV background $J(\nu, z)$ and the photoionisation rate $\Gamma_{\gamma \hi} (z)$ of \citet{hmr}.
At temperatures higher than about $10^5$ K, ionisation by collisional excitation becomes the dominant mechanism. In this regime, the collisional ionisation rate is (\citealt{cen}, \citealt{tom}):
\begin{equation}
\Gamma_{e \hi} = 1.17\times 10^{-10} T^{0.5} e^{-157809.1 /T}
\frac{1}{1+ T_5 ^{0.5}} \textrm{ cm}^3 \textrm{ s}^{-1}.
\end{equation}

\section{Optical depth and flux spectrum}
\label{optical}

The optical depth of the gas is calculated by integrating the relevant quantities (density, temperature and velocity) along a given LOS.
We identify a sightline through the high resolution sphere and
we divide its total length $L$ into $N$ pixels of equal width $\Delta = L/N$.
We choose $N=3000$.
The density and the density--weighted temperature and velocity for each bin $j$
at position $x(j)$ are computed by integrating the relevant quantities along the
LOS \citep{tom}: 
\begin{equation}\label{intdd}
\rho_{\textrm{X}} (j) = a^3 \sum_{\textrm{i}} X(i) \mathcal{W}_{\textrm{ij}},
\end{equation}
\begin{equation}\label{inttt}
\left(\rho T\right)_{\textrm{X}} (j) = a^3 \sum_{\textrm{i}} X(i) \mathcal{W}_{\textrm{ij}} T(i),
\end{equation}
\begin{equation}\label{intvv}
\left(\rho v\right)_{\textrm{X}} (j) = a^3 \sum_{\textrm{i}} X(i) \mathcal{W}_{\textrm{ij}} \left\{ a\dot{x}(i) + \dot{a}\left[ x(i) - x(j) \right] \right\},
\end{equation}
where the sum is over all the particles intersecting the LOS. $a$ is the scale factor and $X(i)$ is the abundance of the species $X$ for particle $i$, assuming ionisation equilibrium. For \lya\ absorption spectra, $X$ is the neutral hydrogen \hi .
The normalised SPH kernel is:
\begin{equation}\label{normk}
\mathcal{W}_{\textrm{ij}} = \frac{mW(r_{\textrm{ij}}/h_{\textrm{i}})}{h_{\textrm{i}} ^3},
\end{equation}
with $W(r_{\textrm{ij}}/h_{\textrm{i}})$ given by
\begin{equation}\label{kernel}
W\left( \mathbf{r}/h \right) = \frac{8}{\pi h^3} \left\{
  \begin{array}{ll}
  1-6\left( \frac{r}{h}\right)^2 + 6\left(\frac{r}{h}\right)^3, & 
     0\leq\frac{r}{h}\leq \frac{1}{2},\\
  2\left( 1-\frac{r}{h}\right)^3, & \frac{1}{2}<\frac{r}{h}\leq 1, \\
  0, & \frac{r}{h} > 1.
  \end{array} \right.
\end{equation}
as used in {\small GADGET I}. $m$ is the mass of the dark matter particles, $r_{ij}$ is the distance between particle $i$ and particle $j$ and $h_i$ the smoothing length of the dark matter particle.

In redshift space, a pixel at velocity $v(k)$ suffers absorption from a pixel at
velocity $v(j)$ by an amount $e^{-\tau (k)}$, where $\tau (k)$ is the contribution to the optical depth of bin $k$ given by bin $j$:
\begin{equation}\label{opticaldepth}
\tau (k) = \frac{1}{\sqrt{\pi}}
\sigma_{\alpha} \frac{c}{V_{\textrm X} (j)} n_{\textrm X} (j) a\Delta \cdot \exp \left\{ -\left[ \frac{v(k) - v(j)}{V_{\textrm X} (j)}\right]^2 \right\}.
\end{equation}

The Doppler width $V_{\textrm X} (j)$ of the species $X$ with mass $m_{\textrm X}$ determines the thermal broadening of the absorption lines. The Doppler width $V_{\textrm X} (j)$ in pixel $j$ is given by:
\begin{equation}
V_{\textrm X} ^2(j) = \frac{2k_{\textrm B} T_{\textrm X} (j)}{m_{\textrm X}},
\end{equation}
where $n_{\textrm X} (j)$ is the numer density, $T_{\textrm X} (j)$ the temperature in pixel $j$ and $c$ the speed of light. The \lya\ cross section for \hi\ is
\begin{equation}
\sigma_{\alpha} = \left( \frac{3\pi \sigma_{\textrm T}}{8}\right)^{1/2} f\lambda_{\textrm o} = 4.45\cdot 10^{-18} \textrm{ cm}^2,
\end{equation}
with $\lambda_{\textrm o} = 1215.6$ \AA\ the rest wavelength of the transition and $\sigma_{\textrm T} = 6.625 \cdot 10^{-25}$ cm$^{2}$ the Thomson cross section. $f= 0.41615$ is the oscillator strength and measures the quantum mechanical departure from the classical harmonic oscillator.

We normalise our spectra by rescaling the mean flux $\langle e^{-\tau_{\hi}} \rangle$ to the mean flux measured in real spectra, $e^{-\tau_{\textrm{eff}}}$. $\tau_{\hi}$ is the \hi\ opacity along our simulated LOS. The effective optical depth $\tau_{\textrm{eff}}$ derived from observations is calculated as a function of redshift by using the relation between redshift and optical depth given by \citet{kim} in Subsection 3.3 of their paper. The effective optical depth $\tau_{\textrm{oss}}$ in our spectra is:
\begin{equation}
\tau_{\textrm{oss}} = - \ln \langle e^{-\tau_{\hi}} \rangle .
\end{equation}

With an iterative procedure we rescale $\tau_{\textrm{oss}}$ to $\tau_{\textrm{eff}}$ until the difference between the two is no larger than a few percent, that is, typically, $\left| \tau_{\textrm{oss}} - \tau_{\textrm{eff}} \right| < 0.005$.
This procedure allows us to make the artificial spectra directly comparable to
the observed ones in each given redshift range.
In addition, it somewhat washes out the dependence of the optical depth on the
normalisation coefficient of the UV ionising background radiation $J_{21}$,
which determines the ionisation fraction of hydrogen and therefore the
absorption of the UV radiation.

\bsp
\label{lastpage}


\begin{thebibliography}{99}
\bibitem[\protect\citeauthoryear{Adelberger et al.}{2003}]{adelb}
Adelberger K.L., Steidel C.C., Shapley A.E. \& Pettini M., 2003, ApJ, 584, 45

\bibitem[\protect\citeauthoryear{Adelberger et al.}{2005}]{adelberger}
Adelberger K.L., Shapley A.E., Steidel C.C., Pettini M., Erb D.K. \& Reddy N.A., 2005, ApJ, 629, 636 (A05)

\bibitem[\protect\citeauthoryear{Aguirre et al.}{2005}]{aguirre2005}
Aguirre A., Schaye J., Hernquist L., Kay S., Springel V., Theuns T., 2005, ApJ, 620, 13

\bibitem[\protect\citeauthoryear{Aguirre et al.}{2001}]{aguirre}
Aguirre A., Hernquist L., Schaye J., Katz N., Weinberg D.H., Gardner J. 2001,
ApJ, 561, 521

\bibitem[\protect\citeauthoryear{Bertone, Stoehr \& White}{2005}]{serena}
Bertone S., Stoehr F. \& White S.D.M., 2005, MNRAS, 359, 1201 (BSW05)

\bibitem[\protect\citeauthoryear{Bolton et al.}{2004}]{bolton}
Bolton J.S., Haehnelt M.G., Viel M., Springel V., 2005, MNRAS, 357, 1178

\bibitem[\protect\citeauthoryear{Cen}{1992}]{cen}
Cen R., 1992, ApJS, 78, 341

\bibitem[\protect\citeauthoryear{Colless et al.}{2001}]{colless}
Colless M. et al., 2001, MNRAS, 328, 1039

\bibitem[\protect\citeauthoryear{Croft et al.}{2002}]{croft}
Croft R.A.C., Weinberg D.H., Bolte M., Burles S., Hernquist L., Katz N., Kirkman D., Tytler D., 2002, ApJ, 581, 20

\bibitem[\protect\citeauthoryear{Croft et al.}{1998}]{croft1998}
Croft R.A.C., Weinberg D.H., Katz N., Hernquist L., 1998, ApJ, 495, 44

\bibitem[\protect\citeauthoryear{Desjacques et al.}{2004}]{vincent}
Desjacques V., Nusser A., Haehnelt M.G., Stoehr F., 2004, MNRAS, 350, 879

\bibitem[\protect\citeauthoryear{Fang et al.}{2005}]{fang}
Fang T., Loeb A., Tytler D., Kirkman D., Suzuki N., 2005, pre--print 
astro--ph/0505182

\bibitem[\protect\citeauthoryear{Haardt, Madau \& Rees}{1999}]{hmr}
Haardt F., Madau P., Rees M.J., 1999, ApJ, 514, 648	
	
\bibitem[\protect\citeauthoryear{Haardt \& Madau}{1996}]{haardt}
Haardt F., Madau P., 1996, ApJ, 461 20	

\bibitem[\protect\citeauthoryear{Hoopes et al.}{2003}]{hoopes}
Hoopes C.G., Heckman T.M., Strickland D.K. \& Howk J.C., 2003, ApJL, 596, 175

\bibitem[\protect\citeauthoryear{Hui et al.}{2001}]{hui}
Hui L., Burles S., Seljak U., Rutledge R.E., Magnier E., Tytler D., 2001, ApJ, 552, 15

\bibitem[\protect\citeauthoryear{Hui \& Gnedin}{1997}]{huignedin}
Hui L., Gnedin N.Y., 1997, MNRAS, 292, 27

\bibitem[\protect\citeauthoryear{Kim et al.}{2002}]{kim02}
Kim T.-S., Carswell R. F., Cristiani S., D'Odorico S., Giallongo E., 2002, MNRAS, 335, 555	

\bibitem[\protect\citeauthoryear{Kim et al.}{2004}]{kim}
Kim T.-S., Viel M., Haehnelt M.G., Carswell R. F., Cristiani S., 2004,
MNRAS, 347, 355

\bibitem[\protect\citeauthoryear{Kollmeier et al.}{2003}]{juna}
Kollmeier J.A., Weinberg D.H., Dav\'e R., Katz N., 2003, ApJ, 594, 75

\bibitem[\protect\citeauthoryear{Lidz et al.}{2005}]{lidz}
Lidz A., Heitmann K., Hui L., Habib S., Rauch M., Sargent W.L.W., 2005, pre--print astro--ph/0505138

\bibitem[\protect\citeauthoryear{Maselli et al.}{2004}]{maselli}
Maselli A., Ferrara A., Bruscoli M., Marri S., Schneider R., 2004, MNRAS, 350, 21

\bibitem[\protect\citeauthoryear{McDonald et al.}{2004}]{mcsdss}
McDonald P., Seljak U., Burles S., Schlegel D.J., Weinberg D.H., Shih D., Schaye J., Schneider D.P., Brinkmann J., Brunner R.J., Fukugita M., 2004, pre--print astro--ph/0405013

\bibitem[\protect\citeauthoryear{McDonald et al.}{2005}]{mc2005}
McDonald P., Seljak U., Cen R., Bode P., Ostriker J.P., 2005, MNRAS, 360, 1471

\bibitem[\protect\citeauthoryear{McDonald et al.}{2000}]{mcd}
McDonald P., Miralda--Escud\'e J., Rauch M., Sargent W.L.W., Barlow T.A., Cen R., Ostriker J.P., 2000, ApJ, 543, 1

\bibitem[\protect\citeauthoryear{Ostriker \& McKee}{1988}]{omk}
Ostriker J.P. \& McKee C.F., 1988, Rev.Mod.Phys. 60, 1

\bibitem[\protect\citeauthoryear{Rupke et al.}{2005}]{rupke}
Rupke D.S., Veilleux S., Sanders D.B., 2005, pre--print astro--ph/0507037.

\bibitem[\protect\citeauthoryear{Schaye et al.}{2000}]{joop}
Schaye J., Theuns T., Rauch M., Efstathiou G., Sargent W.L.W., 2000, MNRAS,
318, 817

\bibitem[\protect\citeauthoryear{Shapley et al.}{2003}]{shapley}
Shapley A.E., Steidel C.C., Pettini M. \& Adelberger K.L., 2003, ApJ, 588, 65

\bibitem[\protect\citeauthoryear{Spergel et al.}{2003}]{spergel}
Spergel D.N., Verde L., Peiris H.V., Komatsu E., Nolta M.R., Bennett, C.L.,
Halpern M., Hinshaw G., Jarosik N., Kogut A., Limon M., Meyer S.S., Page L.,
Tucker G.S., Weiland J.L., Wollack E., Wright E.L., 2003, ApJS, 148, 175

\bibitem[\protect\citeauthoryear{Springel \& Hernquist}{2003}]{sh}
Springel V., Hernquist L. 2003, MNRAS, 339, 312

\bibitem[\protect\citeauthoryear{Springel et al.}{2001a}]{gadget}
Springel V., Yoshida N. \& White S.D.M., 2001a, New Astronomy, 6, 79

\bibitem[\protect\citeauthoryear{Springel et al.}{2001b}]{semi}
Springel V., White S.D.M., Tormen G. \& Kauffmann G., 2001b, MNRAS, 328, 726

\bibitem[\protect\citeauthoryear{Stoehr}{2003}]{felix}
Stoehr F., 2003, PhD Thesis, Ludwig Maximilian Universit\"at, M\"unchen

\bibitem[\protect\citeauthoryear{Strickland \& Stevens}{2000}]{ses}
Strickland D.K. \& Stevens I.R., 2000, MNRAS, 314, 511

\bibitem[\protect\citeauthoryear{Tegmark et al.}{2004}]{tegmark}
Tegmark M. et al, 2004, ApJ, 606, 702

\bibitem[\protect\citeauthoryear{Theuns et al.}{1998}]{tom}
Theuns T., Leonard A., Efstathiou G., Pearce F.R., Thomas P.A., 1998, MNRAS,
301, 478

\bibitem[\protect\citeauthoryear{Theuns et al.}{2002}]{tv}
Theuns T., Viel M., Kay S., Schaye J., Carswell R.F., Tzanavaris P. 2002,
ApJL, 578, 5

\bibitem[\protect\citeauthoryear{Viana, Nichol \& Liddle}{2002}]{viana}
Viana P.T.P., Nichol R.C., Liddle A.R., 2002, ApJ, 569, 75

\bibitem[\protect\citeauthoryear{Viel et al.}{2004a}]{vielcmb}
Viel M., Weller J. \& Haehnelt M.G., 2004a, MNRAS, 355, 23

\bibitem[\protect\citeauthoryear{Viel et al.}{2004b}]{viel2004}
Viel M., Haehnelt M.G. \& Springel V., 2004b, MNRAS, 354, 684

\bibitem[\protect\citeauthoryear{Viel et al.}{2004c}]{viel}
Viel M., Haehnelt M.G., Carswell R.F., Kim T.--S., 2004c, MNRAS, 349, 33

\bibitem[\protect\citeauthoryear{White \& Frenk}{1991}]{wf}
White S.D.M., Frenk C.S., 1991, ApJ, 379, 52
\end{thebibliography}
\end{document}